\documentclass{article}
\PassOptionsToPackage{numbers, sort&compress}{natbib}

\usepackage[preprint]{neurips_2026}
\usepackage{microtype}
\usepackage{graphicx}
\usepackage{subcaption}
\usepackage{multirow}
\usepackage{booktabs} 
\usepackage{amsmath}
\usepackage{array}
\usepackage{makecell}
\usepackage{wrapfig}
\usepackage{amssymb}

\usepackage{algorithm}
\usepackage{algorithmic}

\usepackage[dvipsnames]{xcolor}
\definecolor{commentpurple}{RGB}{232, 166, 189}
\definecolor{commentblue}{RGB}{127, 127, 201}    
\definecolor{commentgreen}{RGB}{100, 160, 120}  

\definecolor{codeblue}{RGB}{0, 0, 255}        
\definecolor{codegreen}{RGB}{20, 130, 70}       
\usepackage{hyperref}

\usepackage{amsmath}
\usepackage{amssymb}
\usepackage{mathtools}
\usepackage{amsthm}

\theoremstyle{plain}

\theoremstyle{definition}

\theoremstyle{remark}

\usepackage[utf8]{inputenc} 
\usepackage[T1]{fontenc}    
\usepackage{hyperref}       
\usepackage{url}            
\usepackage{booktabs}       
\usepackage{amsfonts}       
\usepackage{nicefrac}       
\usepackage{microtype}      
\usepackage{xcolor}         

\title{Diffusion Editing with Soft Mask: Pixel Level Redo of Image and Video with Adjustable Strength}

\author{%
  Candi Zheng\textsuperscript{1,*,\textdagger} \And Yuan Lan\textsuperscript{2,*}
}

\begin{document}

\maketitle
\begingroup
\renewcommand{\thefootnote}{\textdagger}
\footnotetext{Corresponding author: \texttt{czhengac@connect.ust.hk}\\
  \hspace*{1.8em}\hbox to 0pt{\hss\textsuperscript{*}}Equal contribution\\
  \hspace*{1.8em}\hbox to 0pt{\hss\textsuperscript{1}}Department of Mathematics, Hong Kong University of Science and Technology\\
  \hspace*{1.8em}\hbox to 0pt{\hss\textsuperscript{2}}Independent Researcher}
\endgroup

\begin{abstract}
Diffusion models with prompt and reference image-guided editing have seen rapid progress, yet they remain too coarse for pixel-level control. One promising direction is to incorporate a soft mask that specifies spatially varying edit strengths—but training such fine-grained control demands expensive pixel-wise annotations, while existing zero-shot methods often yield unsatisfactory results. We introduce \textbf{SoftPaint}, a new zero-shot sampling method that leverages soft masks to enable a continuous spectrum of edits, from fully preserving the original content to completely re-synthesizing the masked region. Going beyond zero-shot inpainting methods, we design a Langevin-iteration-based sampler that respects per-pixel soft mask strengths, which applies universally to image and video diffusion models, enabling tasks such as video editing. The method is gradient-free, memory-efficient, and achieves smooth, pixel-level edits across multiple image and video backbones.
\end{abstract}

\begin{figure}[h]
\begin{center}
\begin{minipage}{0.3\columnwidth}
    \centering
    \includegraphics[width=\linewidth]{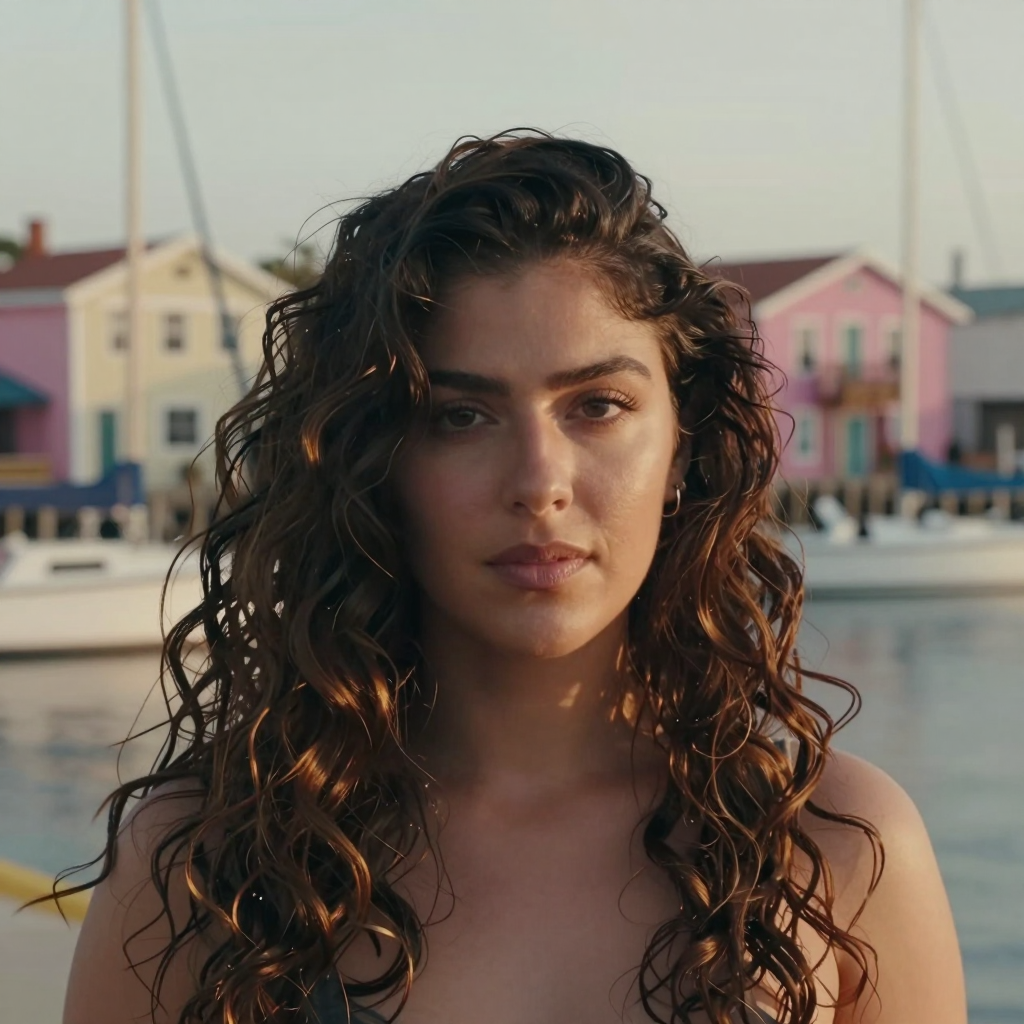}\\
    \small Reference Image
\end{minipage}
\hfill
\begin{minipage}{0.3\columnwidth}
    \centering
    \includegraphics[width=\linewidth]{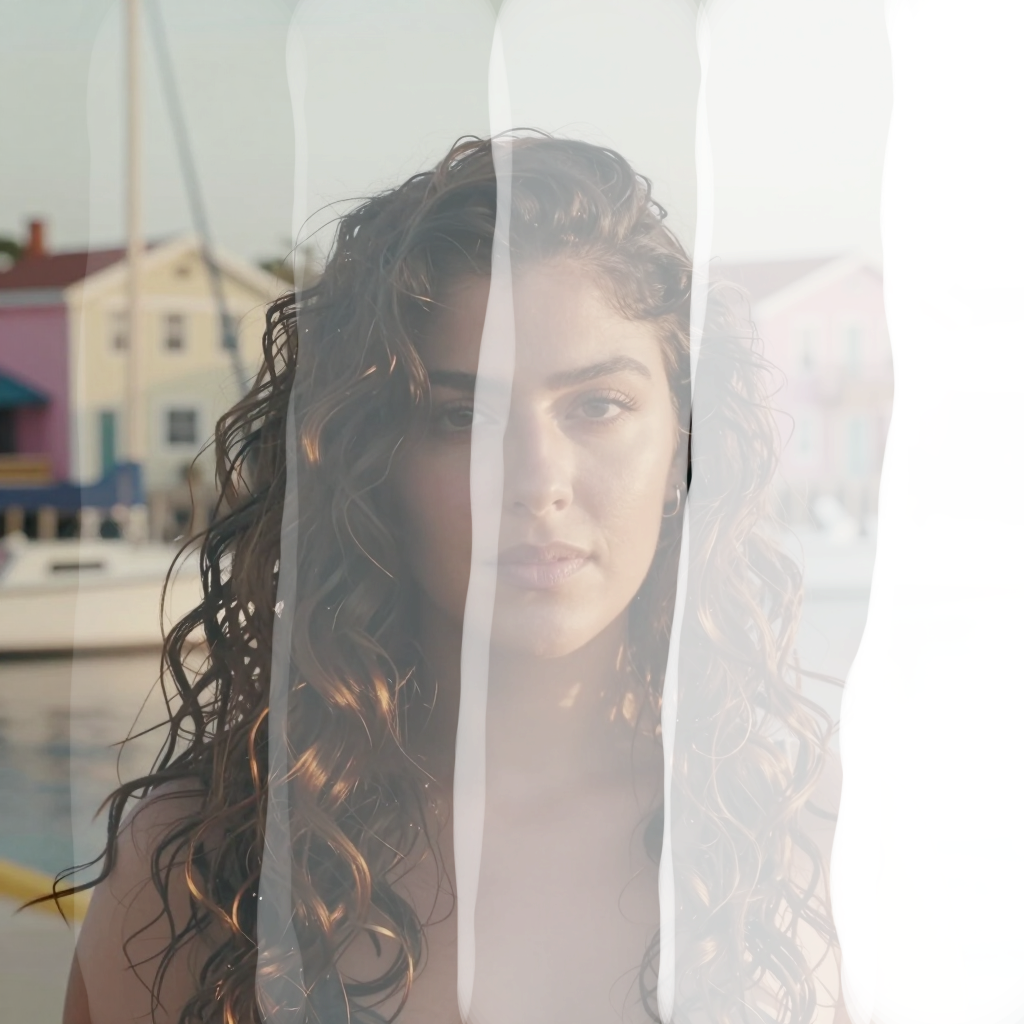}\\
    \small Editing mask (opacity $\propto$ redo)
\end{minipage}
\hfill
\begin{minipage}{0.3\columnwidth}
    \centering
    \includegraphics[width=\linewidth]{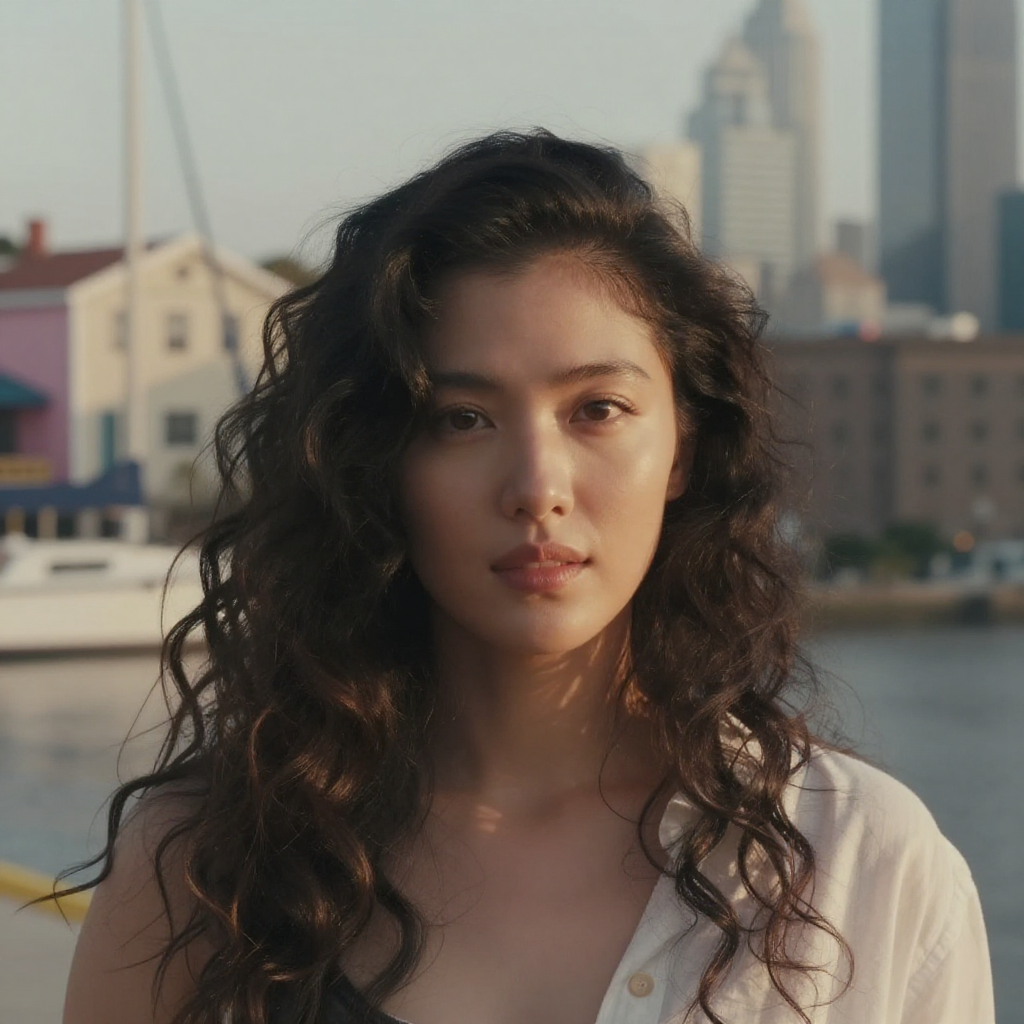}\\
    \small Output
\end{minipage}
\end{center}
\caption{\textbf{Soft Image editing example}. SoftPaint takes a reference image \(\mathbf{x}_0\) (left) and a \emph{continuous-opacity} editing mask (middle), where the mask opacity encodes \emph{redo strength}: more opaque pixels correspond to stronger redo (weaker reference tether), while more transparent pixels keep the sample closer to \(\mathbf{x}_0\) (stronger tether). This yields spatially graded edits within a single generation (right). Text prompt: ``Chinese female with thick wavy black hair, white shirt, night city skyline behind, cinematic close-up.''} \label{fig:soft_example}
\end{figure}

\section{Introduction}

Denoising Diffusion Probabilistic Models (DDPMs) \citep{SohlDickstein2015DeepUL, Song2019GenerativeMB, Song2020ScoreBasedGM, Ho2020DenoisingDP, Rombach2021HighResolutionIS, betker2023improving} have emerged as powerful generative frameworks that produce high-quality outputs through iterative denoising. Subsequent advances in ODE-based deterministic samplers \citep{Karras2022ElucidatingTD, Lu2022DPMSolverFS, Zhao2023UniPCAU}, as well as equivalent rectified flow models \citep{lipman2022flow, liu2022flow, gao2025diffusion}, have dramatically improved the efficiency of DDPMs, reducing the sampling steps from hundreds to dozens. Recent unified generation-and-editing models, such as FLUX.1 Kontext \citep{labs2025flux1kontextflowmatching} and Qwen-Image/Qwen-Image-Edit \citep{wu2025qwenimagetechnicalreport}, further demonstrate how these backbones now support high-quality instruction-based image editing in addition to image synthesis. Together, these innovations have broadened the scope and quality of generative visual art.

Consider a digital artist or a casual user who wishes to edit an image or video. They often have a reference $\mathbf{x}_0$ and want to modify specific parts while preserving others. While text prompts can describe the desired change, they are often too coarse to precisely specify the exact region to be modified. Binary masks, as used in standard inpainting, improve spatial control, but they still enforce a hard decision between ``preserve exactly'' and ``change completely.'' In practice, however, user intent is rarely binary; they often desire a \emph{continuous control of redo strength}---for example, slightly broadening an original smile, or turning the original sky into a dramatic sunset while preserving its rough layout. A natural way to express such intent is through a soft mask, whose values specify how strongly each spatial location should adhere to the reference image. Figure~\ref{fig:soft_example} shows how a single generation with such a continuous-opacity mask yields a spatially graded edit. Furthermore, with the rapid release of diverse open-source models (e.g., Stable Diffusion, FLUX, Qwen-Image) from various organizations, users need a universal editing tool that works \emph{out-of-the-box} for any model, without waiting for model-specific fine-tuning or training specialized adapters.

\begin{figure}[t]
\centering
\includegraphics[width=1\columnwidth]{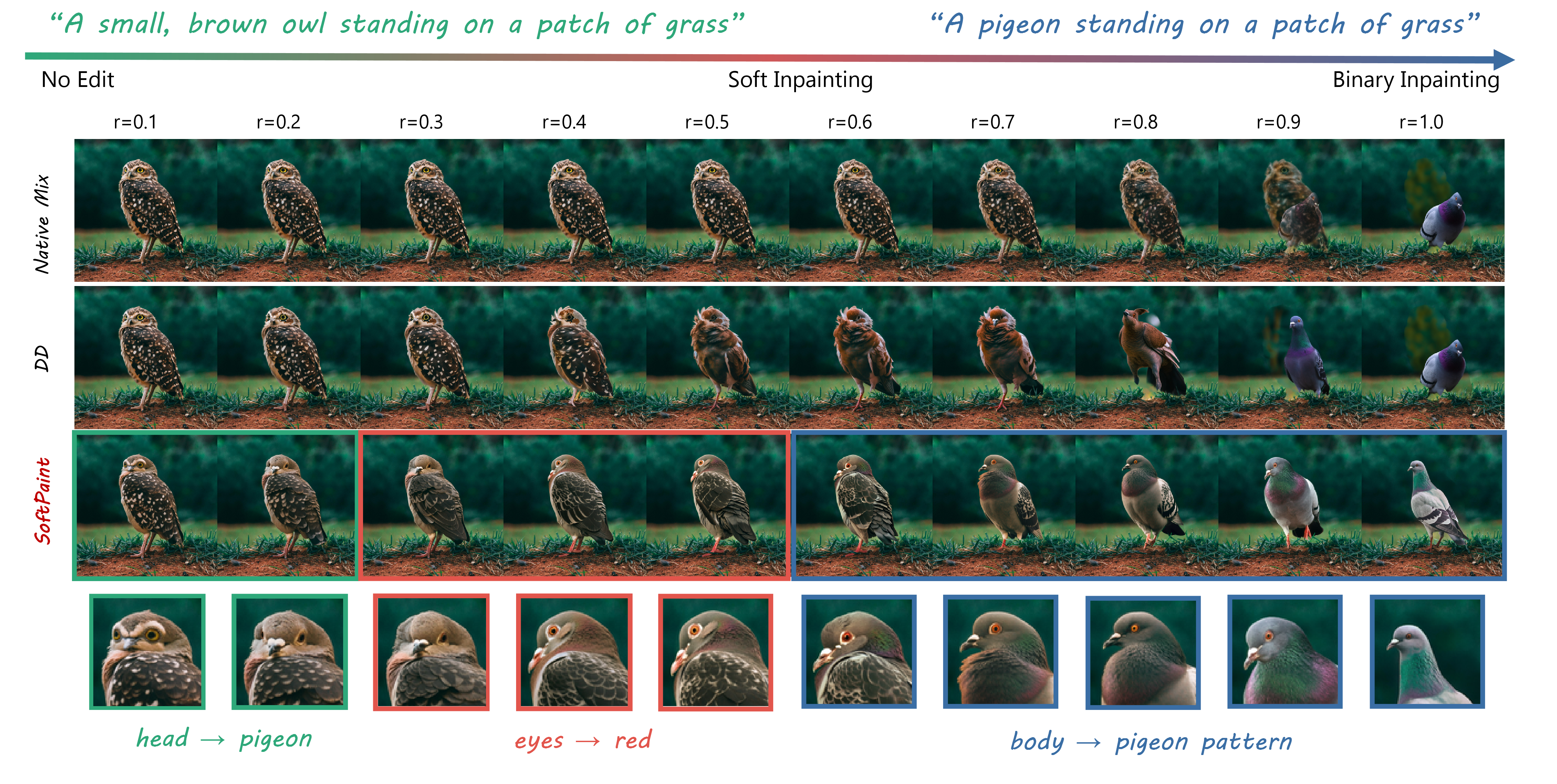}
\caption{\textbf{Fine-grained editing control: $r$ tunes the balance between preservation and redo.}}
\label{fig:img_vis}
\end{figure}

This practical need for \emph{universal, fine-grained control} leads to a fundamental technical challenge: Since diffusion models are trained primarily for whole-image generation, their global denoising mechanism does not naturally support such spatially varying, soft redo. The key question is:
\begin{center}
\emph{Given a diffusion prior $p(\mathbf{x})$, how to sample from a reference-guided distribution $p(\mathbf{x}\mid \mathbf{x}_0, \mathbf{r})$?}
\end{center}
Here \(\mathbf{x}_0\) is a given reference image and \(\mathbf{r}\in[0,1]^d\) is a continuous soft mask: \(\mathbf{r}=0\) means \emph{no redo} (reconstruct the reference), while \(\mathbf{r}=1\) means \emph{full redo} (ignore the reference and recover free generation). Figure~\ref{fig:img_vis} illustrates how a single scalar $r$ shifts this balance between preservation and redo. This formulation represents a \emph{generalized inpainting} problem, where binary inpainting is just a special case with \(\mathbf{r}\in\{0,1\}\).

Current approaches to conditional sampling often face significant trade-offs. Training-based methods \citep{zhang2023adding, mayet2024td, zhuang2024task} are effective for binary inpainting, but extending them to soft redo is much more difficult: preparing supervision for continuous per-pixel redo strengths is expensive, typically requires enormous teams of expert human labelers. On the other hand, training-free methods often either rely on heavy gradient computations (backpropagating through the U-Net), leading to prohibitive GPU memory usage for large production models, or suffer from inferior performance.

In this work, we propose \textbf{SoftPaint}, a pixel-level redo sampler that is \textbf{model-agnostic} and \textbf{training-free}, based on Langevin Dynamics Monte Carlo. SoftPaint supports flow models, as well as ODE- and SDE-based diffusion models. It introduces two core innovations: (1) the \textit{Unified Redo (UR) Score}, a unified guided score for the entire image that combines the diffusion prior score with an analytic Gaussian tether to the reference; and (2) \textit{Diffusion-Damped Langevin (DDL)}, a diffusion-damped Langevin solver that improves stability under large step sizes while keeping the sampling algorithm simple. Experiments confirm that SoftPaint produces high-quality image and video inpainting, while enabling fine-grained redo control for general editing. By avoiding gradient computations, SoftPaint is memory-efficient and deployment-friendly, empowering users to perform soft redo on consumer hardware across diverse diffusion architectures.

\section{From Binary Mask to Soft Mask}
\label{sec:problem-formulation}

Inpainting in diffusion models traditionally relies on binary masks to distinguish between known and unknown regions. 
However, practical editing scenarios often require finer control over the degree of preservation for each pixel. 
This motivates the transition from binary inpainting to \emph{soft inpainting}, where a continuous soft mask controls the redo strength. 
In this section, we formalize both settings and highlight the non-trivial challenges that arise when extending binary masks to continuous ones.

\paragraph{Vanilla Diffusion Inpainting with Binary Masks}
\label{subsubsec:binary-inpainting}

Given an original image $\mathbf{x}_0 \in \mathbb{R}^{H\times W\times 3}$ and a binary mask $\mathbf{m} \in \{0,1\}^{H\times W}$ where $m_{ij}=1$ indicates known pixels and $m_{ij}=0$ indicates missing pixels, diffusion-based inpainting aims to sample from the conditional distribution $p(\mathbf{x} \mid \mathbf{x}_0 \odot \mathbf{m})$. A vanilla approach \cite{song2019generative, lugmayr2022repaint} uses a pre-trained unconditional diffusion model and modifies the reverse process by blending the model sample $\mathbf{x}_t^{\text{model}}$ with the noised reference $\mathbf{x}_t^{\text{ref}}$ at each time step $t$, where $\mathbf{x}_t^{\text{ref}}$ is obtained by forward-diffusing the reference image, i.e., $\mathbf{x}_t^{\text{ref}}=\sqrt{\bar{\alpha}_t}\mathbf{x}_0+\sqrt{1-\bar{\alpha}_t}\boldsymbol{\epsilon}$ with $\boldsymbol{\epsilon}\sim\mathcal{N}(\mathbf{0},\mathbf{I})$:
\begin{equation}
\label{eq:binary-blending}
\hat{\mathbf{x}}_0^{\text{replace}} = \mathbf{m} \odot \mathbf{x}_0^{\text{ref}} + (1 - \mathbf{m}) \odot \hat{\mathbf{x}}_0^{\text{model}}(x_t) .
\end{equation}
This \textit{hard replacement} ensures exact reconstruction of known pixels while freely generating missing regions. Although plausible for binary masks, this method does not naturally extend to continuous masks.

\paragraph{The Challenge of Continuous Masks}
\label{subsubsec:challenge-continuous}

Extending binary masks to a continuous soft mask $\mathbf{r} \in [0,1]^{H\times W}$, where $r_{ij}$ controls the redo strength, is non-trivial. Naively generalizing Eq.~\eqref{eq:binary-blending} as
\begin{equation}
\label{eq:naive-clean-blending}
\hat{\mathbf{x}}_0^{\text{naive}} = (1 - \mathbf{r}) \odot \mathbf{x}^{\text{ref}}_0 + \mathbf{r} \odot \hat{\mathbf{x}}_0^{\text{model}}(x_t),
\end{equation}

Assume $\hat{\mathbf{x}}_0^{\text{model}}$ and $\mathbf{x}^{\text{ref}}_0$ are meaningfully different.
Then their naive combination mixes inconsistent clean images, producing blur (low likelihood under the diffusion prior).
To avoid this low-likelihood outcome, the model tends to reduce the mismatch and push $\hat{\mathbf{x}}_0^{\text{model}} \approx \mathbf{x}^{\text{ref}}_0$.
As a result, outputs stay too close to the reference, so intermediate $\mathbf{r}$ values fail to produce proportional redo.

Instead of naive blending, transitioning to soft inpainting requires a more principled formulation of the target distribution and a corresponding sampling algorithm compatible with diffusion dynamics.

\paragraph{Desired Properties of a Soft Inpainting Formulation}
\label{subsubsec:desired-properties}

To address this challenge, we define a well-behaved target distribution $q(\mathbf{x} \mid \mathbf{x}^{\text{ref}}_0, \mathbf{r})$ parameterized by the continuous soft mask $\mathbf{r} \in [0,1]^{H\times W}$, which must satisfy the following properties:

\begin{enumerate}
    \item \textbf{Well-defined distribution}: $q(\mathbf{x}|\mathbf{x}^{\text{ref}}_0,\mathbf{r})$ must be a valid probability density function.
    
    \item \textbf{Boundary conditions}: At extreme values, $q$ must recover standard behaviors: 
    $\mathbf{r}=0$ yields perfect reconstruction $q(\mathbf{x} \mid \mathbf{x}^{\text{ref}}_0, \mathbf{r}=0) = \delta(\mathbf{x} - \mathbf{x}^{\text{ref}}_0)$, while 
    $\mathbf{r}=1$ reduces to free generation $q(\mathbf{x} \mid \mathbf{x}^{\text{ref}}_0, \mathbf{r}=1) = p(\mathbf{x})$.
    
    \item \textbf{Smooth interpolation}: For intermediate $\mathbf{r}$, $q$ must continuously and smoothly blend between reconstruction and generation, with pixel-wise control.
    
    \item \textbf{Samplability}: $q$ should be tractable to sample using a pre-trained unconditional diffusion model without retraining.
\end{enumerate}

The transition from binary to soft inpainting is \textbf{highly non-trivial} because simple pixel-space interpolation does not yield a valid probability distribution. SoftPaint addresses this by introducing the \textbf{Unified Redo (UR) Score}, derived from a target distribution that satisfies above four properties and enables efficient sampling via a novel Langevin-based algorithm.

\section{Methodology}
\label{sec:method}
\subsection{Unified Redo (UR) Score}
\label{subsec:ur-score}

Section~\ref{sec:problem-formulation} showed that a static blending weight $\mathbf{r}$ is too limited for continuous soft inpainting. We want the reference strength to change \emph{smoothly} over time: early in sampling, the model should stay close to the reference to preserve coherence, while later it should gain more freedom to redo. We also allow the blending weight to \emph{extrapolate} (exceed $1$), so it can go beyond simple interpolation and actively strengthen the influence of the reference image. This extrapolation helps preserve structure and improve coherence.

We propose to replace the static weight $\mathbf{r}$ with a \emph{time- and pixel-dependent} reference blending weight $\boldsymbol{\gamma}(t,\mathbf{r})\in[0,+\infty)^{H\times W}$. The desired clean-image estimate at time $t$ then becomes
\begin{equation}
\label{eq:target_x0}
\hat{\mathbf{x}}_0^{\text{target}}(\mathbf{x}_t) = \boldsymbol{\gamma}(t,\mathbf{r}) \odot \mathbf{x}_0^{\text{ref}} + \bigl(\mathbf{1}-\boldsymbol{\gamma}(t,\mathbf{r})\bigr) \odot \hat{\mathbf{x}}_0^{\text{model}}(\mathbf{x}_t).
\end{equation}
Our goal in this section is: (i) to translate this clean-image target into a tractable sampling distribution over $\mathbf{x}_t$, and (ii) to design the function $\boldsymbol{\gamma}(t,\mathbf{r})$ such that it smoothly interpolates between free generation ($\boldsymbol{\gamma}=\mathbf{0}$) and exact reconstruction ($\boldsymbol{\gamma}$ is large) while providing intuitive control via the user soft mask $\mathbf{r}$.

\paragraph{Probabilistic Reformulation.}
In diffusion models, guiding the sample $\mathbf{x}_t$ so that its predicted clean image follows \eqref{eq:target_x0} is equivalent to sampling $\mathbf{x}_t$ from a distribution that balances two forces: one pulling toward the reference $\mathbf{x}_0^{\text{ref}}$, and one pushing toward the model's own prior. Because the guidance strength $\gamma_t^{(i)}$ can differ across pixels in soft inpainting scenario, we express this balance at the per-pixel level. Concretely, the target distribution corresponding to \eqref{eq:target_x0} over $\mathbf{x}_t$ is a product of pixel-wise geometric mixtures:
\begin{equation}
\label{eq:pixelwise_mixture}
\pi(\mathbf{x}_t) \propto \prod_i \Bigl[ 
\underbrace{p\bigl(x_t^{(i)}\mid x_0^{(i)}\bigr)^{\gamma^{(i)}_t}}_{\text{reference factor}} 
\cdot 
\underbrace{p^{\text{model}}\bigl(x_t^{(i)}\mid x_t^{(-i)}\bigr)^{1-\gamma^{(i)}_t}}_{\text{model factor}} 
\Bigr],
\end{equation}
where $p(x_t^{(i)}\mid x_0^{(i)}) = \mathcal{N}\bigl(x_t^{(i)}\mid \sqrt{\bar{\alpha}_t}x_0^{(i)}, 1-\bar{\alpha}_t\bigr)$ is the forward diffusion kernel (identical across pixels), and $p^{\text{model}}(x_t^{(i)}\mid x_t^{(-i)})$ denotes the pretrained model's implicit conditional for the $i$-th pixel given all other pixels denoted by $x_t^{(-i)}$ . The exponents $\gamma^{(i)}_t$ and $1-\gamma^{(i)}_t$ control the relative influence of the reference and the model at pixel $i$ and time $t$.

At the extremes, \eqref{eq:pixelwise_mixture} recovers the desired boundary behaviors: when $\gamma_t^{(i)}>1$, the $i$-th pixel is strongly tethered to the exact forward process from $x_0^{(i)}$; when $\gamma_t^{(i)}\approx 0$, the reference factor vanishes and the pixel relies entirely on the model. Intermediate values interpolate between these two regimes. The challenge now reduces to designing $\gamma_t^{(i)}$ such that it faithfully reflects the user's soft inpainting intent encoded in $\mathbf{r}$.

\paragraph{Design of the Reference Blending Weight $\boldsymbol{\gamma}(t,\mathbf{r})$.}

A naive choice would be to set $\gamma_t^{(i)} = 1-r_i$. As discussed in 
Section~\ref{sec:problem-formulation}, this static weighting fails to interpolate smoothly. 
Probabilistically, the issue is that the probability mass of $p(x_t^{(i)}\mid x_0^{(i)})^{1-r_i}$ is concentrated at 
$x_0^{(i)}$ for any constant $r_i<1$ when $t=0$. To achieve 
genuine softness, we must allow the model to sample values near $x_0^{(i)}$ when $t=0$.

Our key insight is that a truly soft constraint should allow the generated pixel to deviate 
from the exact reference in a controlled manner. We realize this by imagining that the reference 
pixel $x_0^{(i)}$ is softened by Gaussian noise with variance $\Sigma^{(i)}$ before being 
diffused. In other words, we replace the sharp likelihood $p(x_t^{(i)}\mid x_0^{(i)})$ with a 
\emph{marginalized} version:
\begin{equation}
\mathbb{E}_{\tilde{x}_0^{(i)}\sim\mathcal{N}(x_0^{(i)},\,\Sigma^{(i)})}\Bigl[ p(x_t^{(i)}\mid \tilde{x}_0^{(i)}) \Bigr].
\end{equation}
When $\Sigma^{(i)}$ is large, this marginal density becomes broad, permitting large deviations; 
as $\Sigma^{(i)}\to 0$, it collapses back to the exact constraint.

A crucial algebraic simplification (derived in Appendix~\ref{app:ur-derivation}) shows that 
this expectation is proportional to the original diffusion kernel raised to a power:
\begin{equation}
\label{eq:beta_t_relation}
\mathbb{E}_{\mathcal{N}(x_0^{(i)},\Sigma^{(i)})}\bigl[ p(x_t^{(i)}\mid \tilde{x}_0^{(i)}) \bigr] 
\propto p(x_t^{(i)}\mid x_0^{(i)})^{\beta_t^{(i)}},
\end{equation}
where
\begin{equation}
\label{eq:beta_t}
\beta_t^{(i)} = \frac{1-\bar{\alpha}_t}{1-\bar{\alpha}_t + \bar{\alpha}_t\,\Sigma^{(i)}}.
\end{equation}
This expression suggests that $\beta_t^{(i)}$ should serve as the base time-dependent weight for the reference term. To further extrapolate the reference strength, analogous to the guidance scale in classifier-free guidance~\citep{Ho2022ClassifierFreeDG}, we multiply it by an additional factor $(1+\lambda)$ with $\lambda\ge 0$, yielding the final exponent
\begin{equation}
\label{eq:gamma_t_final}
\gamma_t^{(i)}(\mathbf{r}) = (1+\lambda)\,\beta_t^{(i)} 
= (1+\lambda)\,\frac{1-\bar{\alpha}_t}{1-\bar{\alpha}_t + \bar{\alpha}_t\,\Sigma^{(i)}(\mathbf{r})},
\end{equation}
The redo strength is controlled by the Gaussian variance $\Sigma$, which we define as a customizable function of the soft mask \(\mathbf{r}\):
\(\boldsymbol{\Sigma}(\mathbf{r};\eta) = (1+2\lambda)\,\frac{1-(1-\mathbf{r})^{\eta}}{(1-\mathbf{r})^{\eta}+\epsilon}\),
where \(\eta>0\) is a hyperparameter, the fraction and exponent are applied element-wise, and \(\epsilon=10^{-4}\). Based on the ablation in Appendix~\ref{app:abl-eta} (Figure~\ref{fig:sigma_choice}), we use \(\eta=2\), which makes \(\mathrm{clip}_t\) approximately linear in \(r\).

\begin{figure*}[t]
\centering
\begin{minipage}[c]{0.49\textwidth}
\centering
\includegraphics[width=\linewidth]{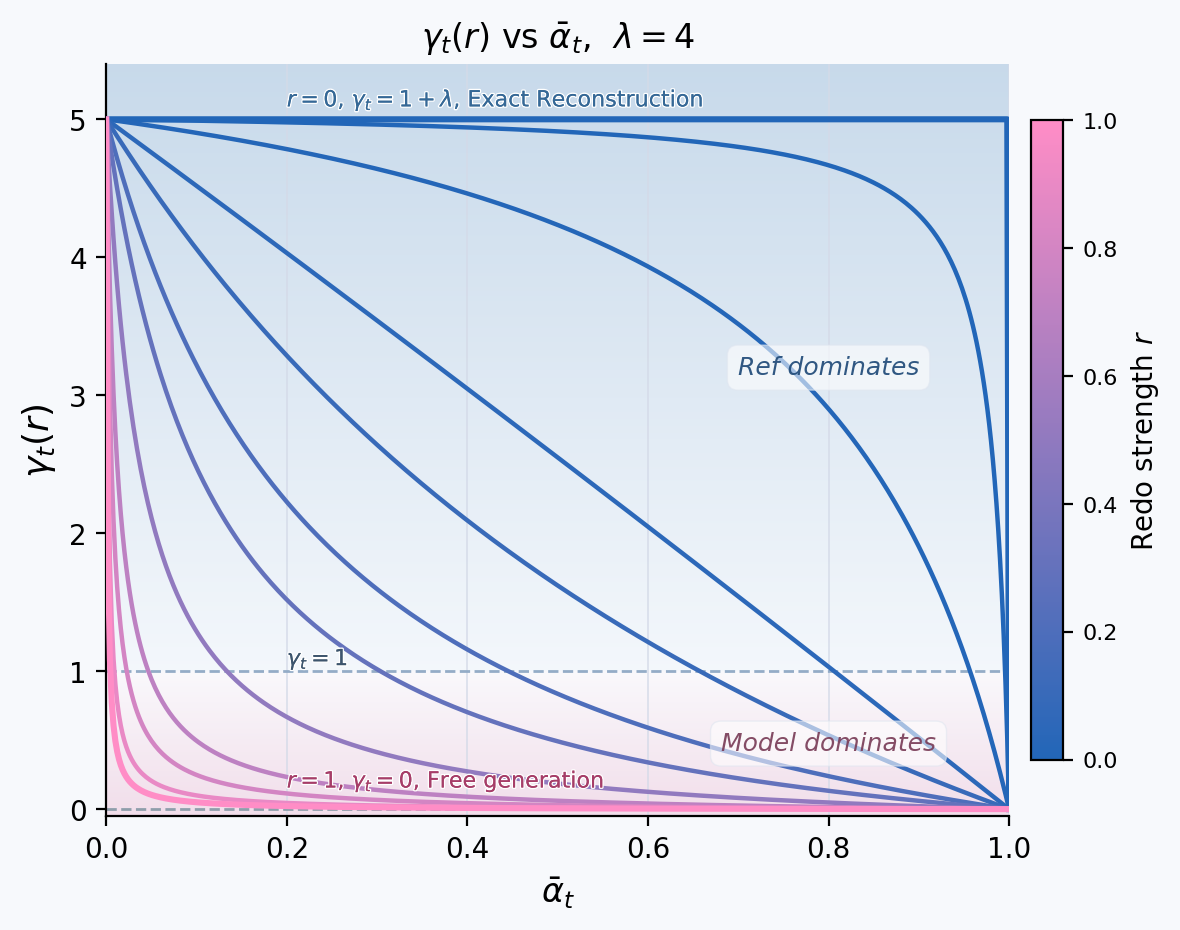}
\captionof{figure}{\textbf{Reference blending weight $\boldsymbol{\gamma}_t(\mathbf{r})$ at $\lambda=4$.}
Curves span $r=0$ (top, exact reconstruction) to $r=1$ (bottom, free generation).}
\label{fig:gamma_curve}
\end{minipage}
\hfill
\begin{minipage}[c]{0.5\textwidth}
\centering
\scriptsize
\setlength{\fboxsep}{4pt}
\colorbox{white}{\fbox{\begin{minipage}{\dimexpr\linewidth-2\fboxsep-2\fboxrule\relax}
\centering
\vspace{2pt}
{\small\textbf{Algorithm } \quad \textit{SoftPaint Sampling}}
\refstepcounter{algorithm}\label{alg:lanpaint_plus_main}
\vspace{4pt}
\hrule
\vspace{4pt}
\begin{algorithmic}[1]
\setlength{\itemsep}{3pt}
\REQUIRE Initial noise $\mathbf{x}_T$, reference image $\mathbf{x}_0^{\text{ref}}$, soft mask $\mathbf{r}$, diffusion schedule $\{\bar{\alpha}_t\}_{t=1}^T$, inner Langevin steps $K$, step size $\Delta\tau$
\FOR{$t=T,\dots,1$}
 \STATE \textcolor{gray}{\textit{// Compute reference blending weight $\boldsymbol{\gamma}_t$ via Eq.~\eqref{eq:gamma_t_final}}}
    \STATE
\textcolor{purple}{$\boldsymbol{\gamma}_t (\mathbf{r}) = (1+\lambda)\,\frac{1-\bar{\alpha}_t}{1-\bar{\alpha}_t + \bar{\alpha}_t\,\Sigma(\mathbf{r})}$}
    \STATE $\mathbf{x}_t^{\text{ref}} \leftarrow \sqrt{\bar{\alpha}_t}\mathbf{x}_0^{\text{ref}} + \sqrt{1-\bar{\alpha}_t}\boldsymbol{\epsilon}$
     \STATE \textcolor{gray}{\textit{// Redo-aware initialization}}
    \STATE \textcolor{codeblue}{$\mathbf{m} \leftarrow \mathbb{I}[\boldsymbol{\gamma}_t > 1]$}
    \STATE \textcolor{codeblue}{$\tilde{\mathbf{x}}_t \leftarrow (\mathbf{1}-\mathbf{m}) \odot \mathbf{x}_t + \mathbf{m} \odot \mathbf{x}_t^{\text{ref}}$}
    \FOR{$k=1,\dots,K$}
     \STATE \textcolor{gray}{\textit{// Compute UR score via Eq.~\eqref{eq:g_lambda}}}
        \STATE \textcolor{purple}{$\mathbf{g}(\tilde{\mathbf{x}}_t,t)
= \boldsymbol{\gamma}_t\,\frac{\sqrt{\bar{\alpha}_t}\mathbf{x}_0^{\text{ref}} -\tilde{\mathbf{x}}_t}{1-\bar{\alpha}_t}
+ \bigl(\mathbf{1}-\boldsymbol{\gamma}_t\bigr)\,\mathbf{s}_{\theta}\bigl(\tilde{\mathbf{x}}_t,t\bigr)$}

         \STATE \textcolor{gray}{\textit{ // Diffusion-damped Langevin step}}
        \STATE \textcolor{codeblue}{Update $\tilde{\mathbf{x}}_t$ via one OU step \eqref{eq:ou_update} with $\mathbf{z}_\tau = \tilde{\mathbf{x}}_t$ }
    \ENDFOR
    \STATE $\mathbf{x}_t \leftarrow \tilde{\mathbf{x}}_t$
    \STATE $\mathbf{x}_{t-1} \leftarrow \text{Denoise}(\mathbf{x}_t, t)$
           \textcolor{gray}{\textit{e.g., // DDIM step}}
\ENDFOR
\STATE \textbf{return} $\mathbf{x}_0$
\end{algorithmic}
\vspace{2pt}
\end{minipage}}}
\end{minipage}
\end{figure*}

Figure~\ref{fig:gamma_curve} shows how the blending weight $\gamma_t$ varies with the noise level $\bar{\alpha}_t$ for different mask values $r$. The dashed line $\gamma_t=1$ separates the reference-dominated regime ($\gamma_t>1$, less redo) from the model-dominated regime ($\gamma_t<1$, more redo). Smaller $r$ keeps $\gamma_t$ larger across the trajectory, so the reference remains more influential, whereas larger $r$ drives $\gamma_t$ more quickly toward zero, allowing greater deviation from the reference.

\paragraph{The Unified Redo Score.}
With $\boldsymbol{\gamma}(t,\mathbf{r})$ defined as above, we can convert the probabilistic mixture \eqref{eq:pixelwise_mixture} into a practical sampling algorithm. Taking the gradient of its log-density yields the score function, we obtain the Unified Redo (UR) Score:
\begin{equation}
\label{eq:g_lambda}
\mathbf{g}(\mathbf{x},t)
= \boldsymbol{\gamma}_t(\mathbf{r}) \odot \underbrace{\frac{\sqrt{\bar{\alpha}_t}\mathbf{x}_0^{\text{ref}} -\mathbf{x}}{1-\bar{\alpha}_t}}_{\text{Score of } p_t(\mathbf{x}\mid \mathbf{x}_0^{\text{ref}})}
+ \bigl(\mathbf{1}-\boldsymbol{\gamma}_t(\mathbf{r})\bigr) \odot \underbrace{\mathbf{s}_{\theta}\bigl(\mathbf{x},t\bigr)}_{\text{Score of }p_t^{\text{model}}(\mathbf{x})}.
\end{equation}

This construction also applies to rectified flow models, since rectified flows and diffusion models are equivalent and convertible into each other.

The UR score $\mathbf{g}(\mathbf{x},t)$ and its underlying distribution $\pi(\mathbf{x}_t)$ fulfill the four properties of Section~\ref{subsubsec:desired-properties}. First, $\pi(\mathbf{x}_t)$ is a proper product of probability densities, ensuring a well-defined target. Second, the boundary conditions are met: $\mathbf{r}=\mathbf{0}$ gives $\Sigma=0$, $\beta_t=1$, and $\gamma_t=1+\lambda$, which for $\lambda=0$ enforces exact reconstruction; $\mathbf{r}=\mathbf{1}$ yields $\Sigma\to\infty$, $\beta_t\to 0$, and $\gamma_t\to 0$, reducing $\mathbf{g}$ to the unconditional model score. Third, intermediate $\mathbf{r}$ smoothly interpolates between these extremes because $\Sigma(\mathbf{r})$ continuously modulates $\beta_t$ and hence $\gamma_t$. Finally, $\mathbf{g}(\mathbf{x},t)$ requires only the pretrained score $\mathbf{s}_{\text{model}}$ and an analytical reference term, making it readily samplable without retraining, as detailed in the next section.

\subsection{Diffusion-Damped Langevin (DDL)}
\label{subsec:ddl}

The UR score $\mathbf{g}(\mathbf{x},t)$ derived in Section~\ref{subsec:ur-score} defines the target distribution of the current noisy sample $\mathbf{x}_t$ via Eq.~\eqref{eq:pixelwise_mixture}. A standard reverse diffusion step, however, is designed to follow the model distribution $p^{\text{model}}$ rather than this new target. We therefore refine $\mathbf{x}_t$ at each diffusion step $t$ by running a short inner-loop sampler driven by $\mathbf{g}$ before taking the denoising step to $t-1$.

We consider the overdamped Langevin dynamics (with drift $\mathbf{g}$ from Eq.~\ref{eq:g_lambda}):
\begin{equation} \label{eq:langevin_main}
d\mathbf{z}_\tau = \mathbf{g}(\mathbf{z}_\tau, t)\,d\tau + \sqrt{2}\,d\mathbf{W}_{\tau},
\end{equation}
where $\tau$ is the inner-loop time, $\mathbf{z}_0 \equiv \mathbf{x}_t$, and $\mathbf{W}_\tau$ is standard Brownian motion. In practice, we perform only a few inner updates at each diffusion step $t$.

A naive Euler discretization of Eq.~\eqref{eq:langevin_main} is unstable under strong guidance~\citep{zheng2024characteristic}, so we use a damped exponential integrator. Let $A_t$ be a damping coefficient and define the residual drift
\[
\mathbf{C}_t(\mathbf{z}) := \mathbf{g}(\mathbf{z},t) + A_t \mathbf{z}.
\]
Then one DDL update is
\begin{equation}\label{eq:ou_update}
\begin{aligned}
\mathbf{z}_{\tau+\Delta\tau}
={}& e^{-A_t\Delta\tau}\mathbf{z}_{\tau}
  + \bigl(1-e^{-A_t\Delta\tau}\bigr)A_t^{-1}\mathbf{C}_t(\mathbf{z}_{\tau}) \\
 & + \sqrt{\frac{1-e^{-2A_t\Delta\tau}}{A_t}}\,\boldsymbol{\xi},
\quad \boldsymbol{\xi}\sim\mathcal{N}(\mathbf{0},\mathbf{I}).
\end{aligned}
\end{equation}
where $\Delta\tau$ is the inner-loop step size and $\boldsymbol{\xi}$ is Gaussian noise. This yields a simple and stable inner-loop sampler. The derivation of Eq.~\eqref{eq:ou_update}, the choice of $A_t$, and the stability analysis are deferred to Appendix~\ref{app:ddl-derivation}.

\paragraph{Redo-aware initialization.}
As a practical acceleration, we optionally warm-start the inner Langevin chain from a redo-aware initialization that biases strongly tethered pixels toward the noised reference. This heuristic does not change the stationary distribution and is described in Appendix~\ref{app:abl-warm-start}. Appendix~\ref{app:binary-inpainting} compares the DDL inner loop with a RePaint-style inner loop under binary masks.

\section{Experiments}
\label{sec:experiments}

We evaluate SoftPaint on two soft-mask editing benchmarks: soft image editing on FlowEdit~\citep{kulikov2025flowedit} (Qwen-Image~\citep{wu2025qwenimagetechnicalreport}), and soft video editing on DAVIS-Edit~\citep{liu-etal-2024-stablev2v} (CogVideoX-5B~\citep{yang2024cogvideox}, Wan2.2~\citep{wan2025}). SoftPaint natively supports \emph{per-pixel} soft masks $\mathbf{r}\!\in\![0,1]^{H\times W}$. We use uniform $r$ within the dataset-provided region for controlled edit-strength sweeps, then evaluate a spatially varying four-zone mask separately in Table~\ref{tab:spatially-varying-mask}. A binary-mask comparison isolating the inner-loop operator appears in Appendix~\ref{app:binary-inpainting}. Hyperparameters are ablated in Appendix~\ref{app:ablations}.

\paragraph{Baselines.} We compare against two zero-shot methods with continuous soft-mask support: (i) \textbf{Differential Diffusion (DD)}~\citep{levin2025differential}, the most relevant training-free per-pixel method; and (ii) \textbf{Naive mixing} ($n{=}0$), which disables SoftPaint's inner Langevin loop, reducing it to per-step RePaint-style blending~\citep{lugmayr2022repaint}. Where applicable we also report Qwen-Image's instruction-tuned editor (Qwen-Image-Edit) as a single reference point.

\paragraph{Metric anchoring.} Soft-mask methods trade edit strength against reference preservation along a continuous curve; the same $r$ corresponds to \emph{different} effective edit strengths for DD and SoftPaint (Fig.\ref{fig:image_curves} left panel). We therefore anchor all comparisons at matched CLIP-Text, a backbone-independent proxy for edit strength, and report other metrics at those anchors.

\subsection{Soft-Mask Image Editing}
\label{subsec:exp-image}

This experiment asks two questions: (i)~Does SoftPaint provide \emph{finer controllability} as well as smooth edit strength? (ii)~At matched controllability, does SoftPaint produce \emph{higher-quality} edits? The FlowEdit benchmark~\citep{kulikov2025flowedit} is well-suited here because its diverse object-centric masks and paired source/target prompts let us decouple ``how much the edit was realized'' (CLIP-Text) from ``how clean the result looks'' (LPIPS, PSNR, CLIP-IQA, FID).

\paragraph{Setup.} All three soft-mask methods (SoftPaint, DD, and Naive Mixing) use the same Qwen-Image backbone with $20$ denoising steps on the FlowEdit benchmark~\citep{kulikov2025flowedit}. SoftPaint uses $n{=}2$, $s{=}0.2$, guidance scale $\lambda{=}5$, where $n$ is the number of inner iterations per diffusion step and $s$ is the inner-loop step size. The soft-mask strength $r \in\!\{0,0.1,\ldots,1.0\}$ is swept jointly across DD, Naive Mixing, and SoftPaint. We also report Qwen-Image-Edit once using its default inference setting. Metrics follow the standard FlowEdit protocol: LPIPS$_{\text{edit}}$ and PSNR$_{\text{edit}}$ for edit quality, PSNR$_{\text{keep}}$ for background preservation, CLIP-Text for edit alignment, CLIP-IQA for no-reference quality, and FID for distributional realism.

\paragraph{Edit controllability.} Figure~\ref{fig:image_curves} (left panel) maps $r$ to CLIP-Text for all three methods on Qwen-Image. SoftPaint shows the smoothest behavior, with CLIP-Text growing nearly linearly as $r$ increases. DD is less sensitive at low mask strength, with CLIP-Text changing more slowly in the weak-edit regime. Naive Mixing remains pinned near the source level (CLIP-T $\approx 29.0$) until $r{>}0.8$, then overshoots to $35.4$ at $r{=}1.0$, indicating that it fails to provide smooth edit-strength control. The right three panels trace the quality trade-off: at matched CLIP-Text, SoftPaint achieves lower LPIPS, higher PSNR, and higher CLIP-IQA, with the gap widening at stronger edits, indicating finer controllability. Beyond training-free baselines, we also compare against Qwen-Image-Edit, a modern unified instruction-based editing model. This Qwen-Image-Edit baseline (CLIP-T $30.2$, LPIPS $0.306$) appears far from both soft-mask curves, indicating worse LPIPS and PSNR; that is, although it achieves the edit goal, it deviates substantially from the given reference image. More apples-to-apples comparison  is in Appendix Table~\ref{tab:exp_image_main}.

\paragraph{Equal-NFE comparison.} SoftPaint uses $20$ outer steps and $n{=}2$ additional inner model evaluations per step, for $60$ network function evaluations (NFE) in total. To test whether extra compute explains the Qwen-Image results, we also run DD for $60$ outer steps. At the weak, medium, and strong matched-CLIP-T anchors, SoftPaint has lower LPIPS$_{\text{edit}}$ and higher PSNR$_{\text{edit}}$ than DD at the same $60$-NFE budget. Appendix Table~\ref{tab:exp_image_equal_nfe} reports all five metrics alongside the original $20$-NFE DD results.

\begin{figure*}[t]
\centering
\includegraphics[width=\textwidth]{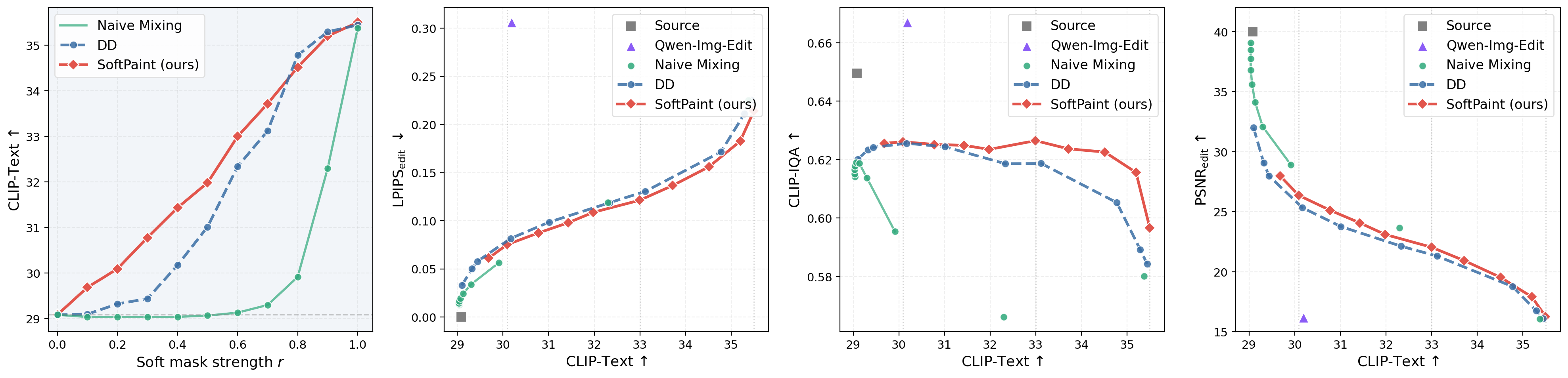}
\caption{\textbf{Image-editing controllability and quality on Qwen-Image (FlowEdit Benchmark).} Left: soft-mask strength $r$ vs.\ CLIP-Text---SoftPaint provides the widest editing range. Right three panels: quality metrics against CLIP-Text as $r$ sweeps from $0.1$ to $1.0$. SoftPaint Pareto-dominates DD; Naive Mixing has near-zero controllability. Source ($r{=}0$, gray square) and Qwen-Image-Edit (purple triangle) are reference points.}
\label{fig:image_curves}
\end{figure*}

\paragraph{Spatially varying redo strengths.} To test nonuniform soft masks directly, we divide each FlowEdit mask into four vertical zones with strengths $r\in\{1,0.75,0.5,0.25\}$. This piecewise-constant mask assigns neighboring regions different redo strengths within the same image. The resulting CLIP-Text scores fall in the medium-edit anchor range ($33.0\pm0.2$). Table~\ref{tab:spatially-varying-mask} shows that SoftPaint improves all four reported quality metrics over DD, including when DD is given the same $60$-NFE budget.

\begin{table}[t]
\centering
\small
\setlength{\tabcolsep}{5pt}
\caption{\textbf{Spatially varying soft-mask editing on Qwen-Image (FlowEdit).} Each image uses four vertical redo-strength zones, $r\in\{1,0.75,0.5,0.25\}$. All methods are evaluated at the medium CLIP-T anchor ($33.0\pm0.2$); the $60$-NFE rows compare equal model-evaluation budgets.}
\label{tab:spatially-varying-mask}
\begin{tabular}{l c c c c c}
\toprule
Method & NFE & PSNR$_{\text{edit}}\uparrow$ & LPIPS$_{\text{edit}}\downarrow$ & FID$\downarrow$ & CLIP-IQA$\uparrow$ \\
\midrule
DD & 20 & 20.3053 & 0.1343 & 102.8441 & 0.6088 \\
DD & 60 & 20.2097 & 0.1349 & 103.4106 & 0.6045 \\
SoftPaint & 60 & \textbf{21.0159} & \textbf{0.1293} & \textbf{100.9059} & \textbf{0.6226} \\
\bottomrule
\end{tabular}
\end{table}

\subsection{Soft-Mask Video Editing}
\label{subsec:exp-video}

\paragraph{Setup.} We run CogVideoX-5B and Wan2.2 at their official default schedules on the DAVIS-Edit ~\citep{liu-etal-2024-stablev2v} benchmark. The exact denoising-step settings are reported in Appendix~\ref{app:ablations}. SoftPaint uses $n{=}2$, $s{=}0.2$ on Wan and CogVideoX. The soft mask strength $r \in\!\{0.1,\ldots,1.0\}$ is swept jointly across the three methods. We use the standard DAVIS-Edit protocol---per-frame PSNR/LPIPS and masked-region M-LPIPS---augmented with VFID for distributional video quality and DOVER for no-reference video aesthetics.

\paragraph{Full-sweep comparison.} Figure~\ref{fig:video_curves} plots the three methods across the full $r$ sweep on both backbones. SoftPaint consistently achieves lower LPIPS, higher PSNR, and lower VFID than DD at matched CLIP-Text, with the gap widening at mid-to-strong edit strengths. Naive Mixing barely shifts CLIP-Text above the source level for $r<0.8$, then overshoots dramatically at $r{=}1.0$ while quality metrics collapse.
Table~\ref{tab:exp_video_main} (Appendix) provides more results compared to DD at matched CLIP-T anchors .

\subsection{Qualitative Comparison}
\label{subsec:exp-qualitative}

Figure~\ref{fig:exp_qualitative} shows representative SoftPaint outputs at $r{=}0.8$ across diverse FlowEdit prompts on Qwen-Image. The unmasked region is preserved verbatim while edits realize the requested semantic change inside the masked region.

\begin{figure*}[t]
\centering
\includegraphics[width=\textwidth]{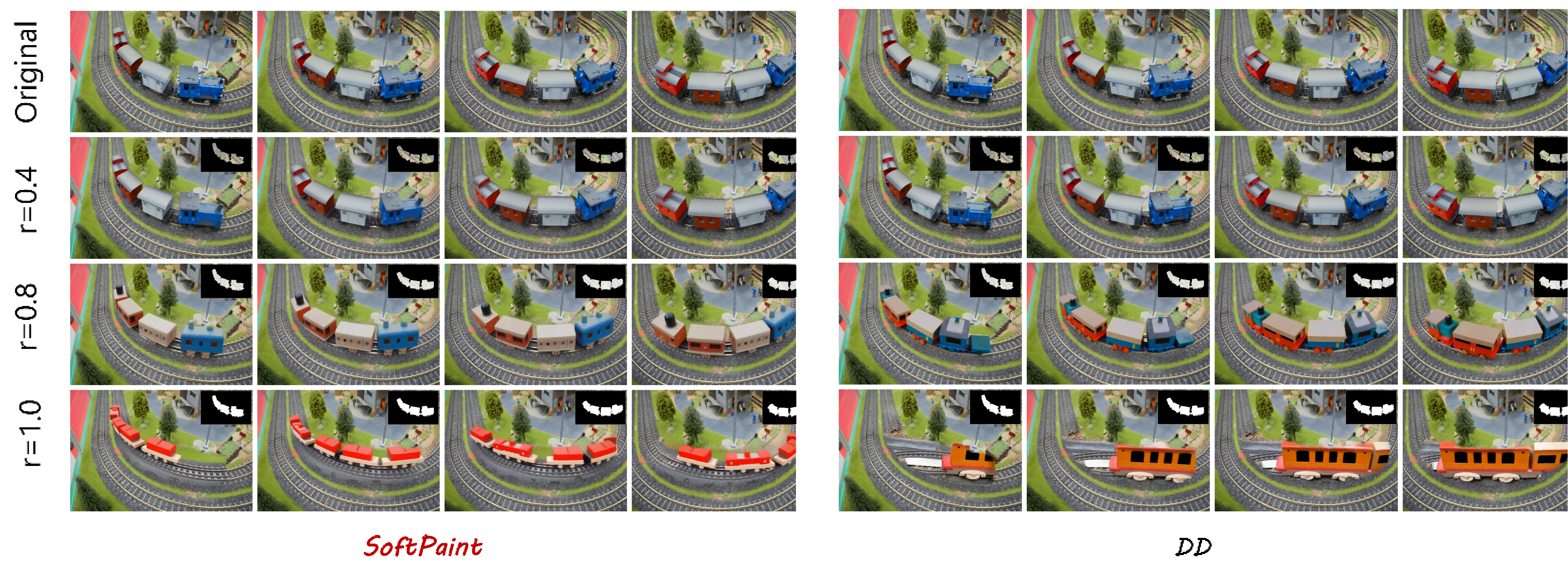}
\caption{\textbf{Soft video editing: ``Model Train'' $\to$ ``Wooden Train''} at three soft-mask strengths $r$ on CogVideoX. Top row: original DAVIS-Edit frames ($r{=}0$, no edit). Subsequent rows: SoftPaint and DD outputs at $r{=}0.4$, $0.8$, $1.0$.}
\label{fig:exp_video_qual}
\end{figure*}

For video, Fig.~\ref{fig:exp_video_qual} compares SoftPaint and DD on a ``Model Train'' $\to$ ``Wooden Train'' edit on CogVideoX as the soft-mask strength $r$ increases. SoftPaint shows a clear coarse-to-fine progression: at $r{=}0.4$, the texture of the train head begins to change while the overall structure is still preserved; at $r{=}0.8$, the original shape is largely maintained but the carriage is already converted to wood; at $r{=}1.0$, the full train is transformed. This behavior is consistent with our design for soft redo strength. By contrast, DD already exhibits missing front-train structure at $r{=}0.8$, and at $r{=}1.0$ produces obvious seams against the surrounding regions, leading to severe artifacts. Additional examples are provided in Appendix~\ref{app:qualitative}.

\begin{figure*}[t]
\centering
\begin{minipage}[c]{0.49\textwidth}
\centering
\includegraphics[width=\linewidth]{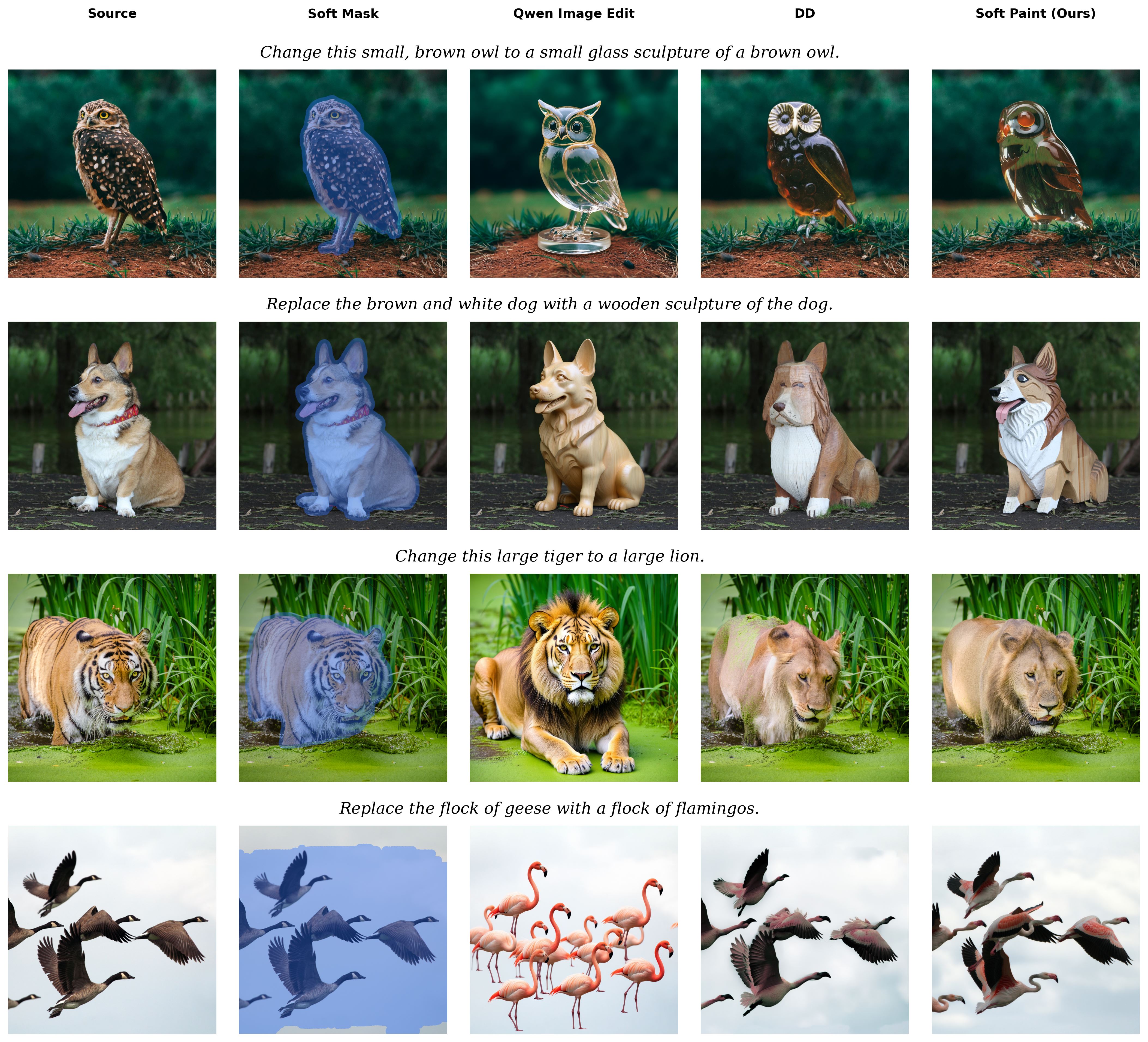}
\setcounter{figure}{4}
\caption{\textbf{Qualitative SoftPaint image edits at $r{=}0.8$.} Each row: source, soft mask, SoftPaint output, prompt. Backbone: Qwen-Image.}
\label{fig:exp_qualitative}
\end{minipage}\hfill
\begin{minipage}[c]{0.49\textwidth}
\centering
\includegraphics[width=0.95\linewidth]{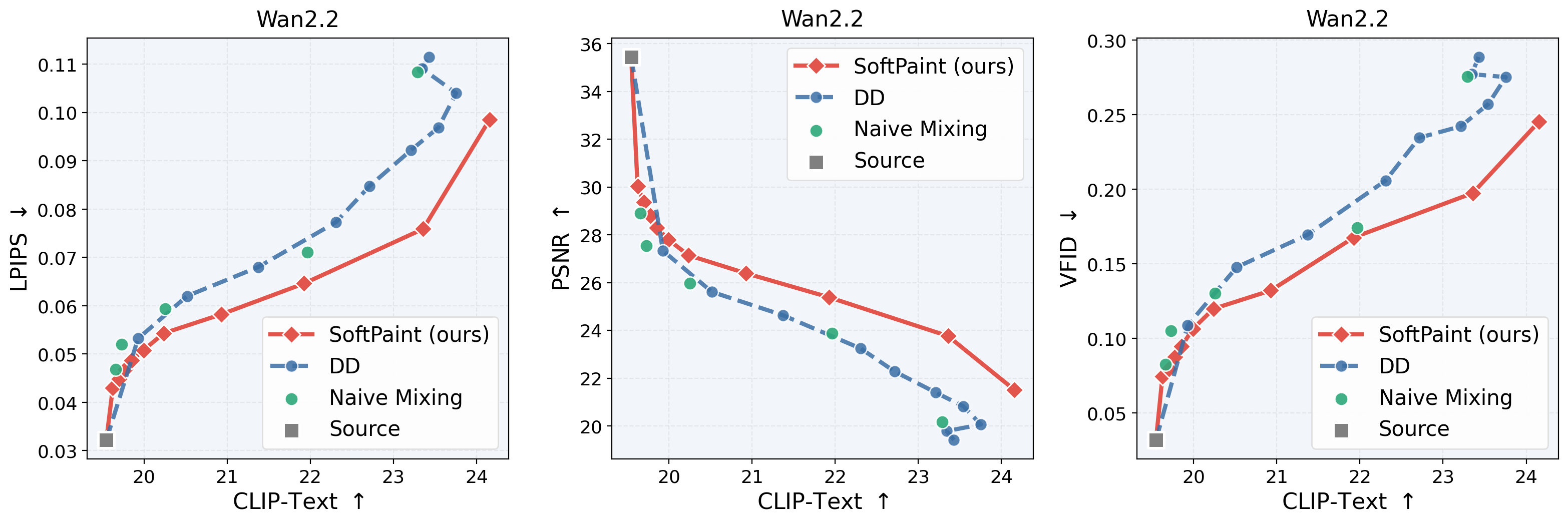}

\vspace{6pt}
\includegraphics[width=0.95\linewidth]{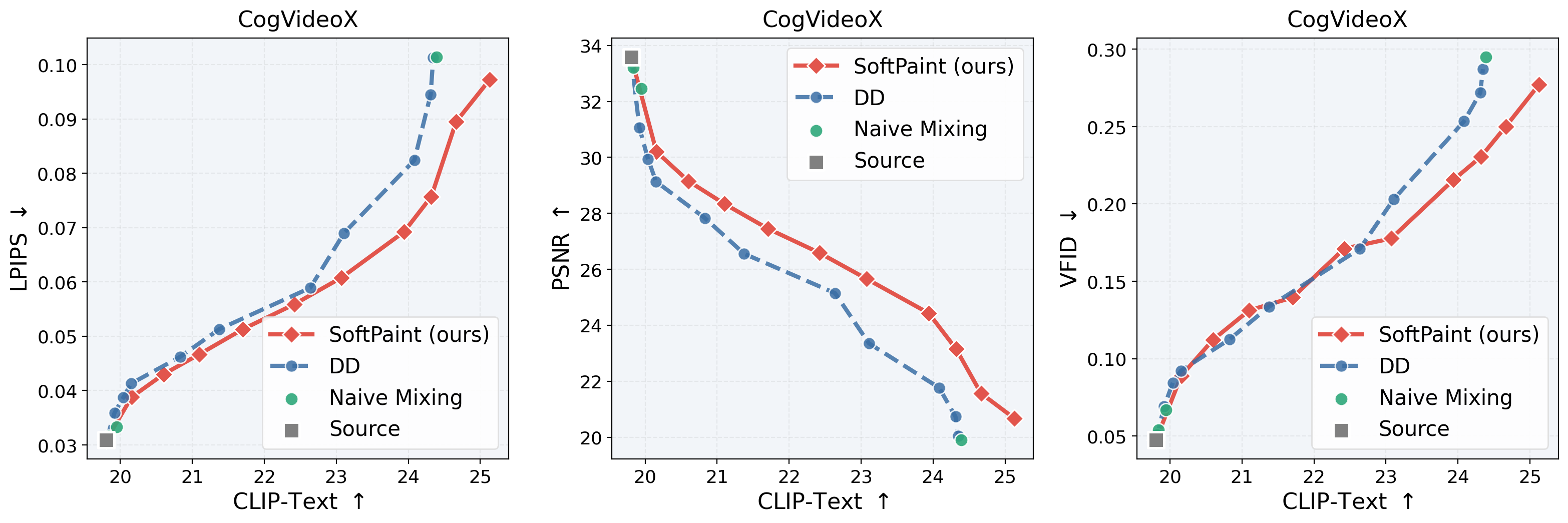}
\setcounter{figure}{5}
\caption{\textbf{Soft video editing on DAVIS-Edit.} Top: Wan2.2. Bottom: CogVideoX. Each panel plots a quality metric against CLIP-Text as the soft-mask strength $r$ is swept. SoftPaint (red) Pareto-dominates DD (blue). Naive Mixing (green circles) has near-zero controllability.}
\label{fig:video_curves}
\end{minipage}
\end{figure*}

\section{Related Works}

\textbf{Diffusion Models and Image/Video Editing.} Diffusion models~\citep{Ho2020DenoisingDP} have become a dominant paradigm for high-quality image and video generation. Early work accelerated the stochastic DDPM sampling process via deterministic ODE flows, notably DDIM~\citep{song2020denoising} and advanced ODE solvers, enabling fewer-step generation. Rectified flow~\citep{lipman2022flow, liu2022flow} further reparameterized the ODE-based diffusion with improved numerical properties, later scaling to large text-to-image and text-to-video models such as Stable Diffusion~3~\citep{esser2024scaling} and FLUX~\citep{blackforest2025flux2}.

Beyond generation, diffusion models have been widely adopted for \emph{image and video editing}, where an input image or video together with a text prompt specifies the desired edit. A large body of work has tackled this problem from different angles: instruction-based editing via fine-tuning on paired data~\citep{brooks2023instructpix2pix}, spatial conditioning through plug-in modules~\citep{zhang2023adding}, image-prompt adaptation with decoupled cross-attention~\citep{ye2023ip}, and multi-task frameworks unifying diverse editing operations~\citep{sheynin2023emu}. More recently, large-scale proprietary models such as FLUX.2 Kontext~\citep{blackforest2025flux2} and Qwen-Image~\citep{wu2025qwenimagetechnicalreport} have integrated reference-image conditioning directly into their training pipelines, supporting multi-reference and text-image-to-image editing with strong semantic preservation.

Despite these advances, existing image and video editing models operate at the \emph{semantic} level and provide limited fine-grained \emph{spatial} control—users cannot specify how strongly each individual pixel should be edited. This gap motivates our approach, which introduces a soft mask mechanism for continuous, pixel-level edit-strength control in a training-free manner.

\textbf{Training-Free Conditional Editing with Diffusion Models.} A line of training-free methods tackles conditional editing by enforcing spatial constraints during the reverse diffusion process. Langevin-based approaches~\citep{zheng2026rethinkingthediffusion} such as RePaint~\citep{lugmayr2022repaint} and LanPaint~\citep{zheng2025lanpaint} perform iterative denoise--renoise under binary inpainting masks~\citep{cornwall2024trainingfree}, while Sequential Monte Carlo methods~\citep{wu2024practical, trippe2022diffusion} and linear-inverse-based techniques~\citep{chung2022improving, chung2022diffusion, grechka2024gradpaint, janati2024divideandconquerposteriorsamplingdenoising, kawar2022denoising, zhang2023towards, rozet2024learning} address similar hard-constraint settings. However, these methods share two key limitations: (i)~they are designed for \emph{binary} masks and do not support continuous-opacity soft masks, and (ii)~many rely on gradient-based optimization or line search, incurring 2--4$\times$ higher GPU memory than standard inference.

Differential Diffusion~\citep{levin2025differential} is a notable exception that enables per-pixel edit-strength control, effectively supporting soft masks in a training-free manner. However, its underlying editing capability is limited: on binary masks it reduces to the naive replace method~\citep{lugmayr2022repaint}, which in turn constrains its soft-mask performance. Our proposed method, SoftPaint, addresses both gaps: it natively supports continuous soft masks with a memory-efficient, gradient-free sampler that is compatible with modern ODE-based diffusion backbones.

\section{Conclusion and Limitations}

We introduced SoftPaint, a training-free and model-agnostic sampler for continuous soft-mask image and video editing. By combining the Unified Redo score with Diffusion-Damped Langevin, SoftPaint enables pixel-level redo control while remaining gradient-free and memory-efficient. Experiments show improved edit controllability and a better quality-preservation trade-off. 

 A practical consideration for SoftPaint is its inference speed. As a training-free method that requires no additional optimization or fine-tuning, SoftPaint can be directly applied to any pretrained diffusion backbone with the same memory footprint as standard generation. The trade-off is that each inner Langevin iteration introduces one extra network evaluation, making the overall generation roughly 
$2\times$--$5\times$ slower than the backbone's standard sampling.  This latency can be a practical limitation for interactive editing and large-scale deployment. However,we believe the zero-training cost and universal compatibility offer a reasonable trade-off; further speed improvements are left to future work.

\paragraph{Broader Impacts} Our work can benefit creative workflows, content restoration, and research on controllable generative modeling by enabling finer image and video edits. However, the same capability could be misused to produce deceptive edits of real scenes or people.

\bibliography{neurips_2026}


\appendix

\section{Additional Tables}

This section provides the detailed matched-CLIP-T tables referenced in the main experiments. Table~\ref{tab:exp_video_main} complements Section~\ref{subsec:exp-video} with anchor-based comparisons on DAVIS-Edit, and Tables~\ref{tab:exp_image_equal_nfe} and~\ref{tab:exp_image_main} complement Section~\ref{subsec:exp-image} with equal-NFE and matched-CLIP-T results on FlowEdit.

\begin{table*}[h!]
\centering\small\setlength{\tabcolsep}{3pt}
\caption{\textbf{Video-editing on DAVIS-Edit at matched CLIP-T.} Anchors are SoftPaint at $r\in\{0.2,0.5,1.0\}$ (Wan2.2) and $r\in\{0.2,0.5,0.9\}$ (CogVideoX). For DD we select the nearest measured $r$ within $\pm0.5$ of each anchor.}
\label{tab:exp_video_main}
\begin{tabular}{l l l c c c c c c}
\toprule
Backbone & Anchor & Method & $r$ used & LPIPS$\downarrow$ & PSNR$\uparrow$ & M-LPIPS$\downarrow$ & VFID$\downarrow$ & DOVER$\uparrow$ \\
\midrule
\multirow{6}{*}{Wan2.2 V2V}
 & \multirow{2}{*}{\shortstack{Weak Edit\\(CLIP-T: 19.7$\pm$0.5)}} & DD        & 0.1 & 0.053 & 27.34 & 0.022 & 0.109 & 54.46 \\
 &  & SoftPaint & 0.2 & \textbf{0.045} & \textbf{29.36} & \textbf{0.015} & \textbf{0.080} & \textbf{55.55} \\
\cmidrule(lr){3-9}
 & \multirow{2}{*}{\shortstack{Medium Edit\\(CLIP-T: 20.0$\pm$0.5)}} & DD        & 0.2 & 0.062 & 25.61 & 0.029 & 0.147 & 54.39 \\
 &  & SoftPaint & 0.5 & \textbf{0.051} & \textbf{27.80} & \textbf{0.020} & \textbf{0.106} & \textbf{55.63} \\
\cmidrule(lr){3-9}
 & \multirow{2}{*}{\shortstack{Strong Edit\\(CLIP-T: 24.2$\pm$0.5)}} & DD        & 0.8 & 0.104 & 20.07 & 0.065 & 0.275 & \textbf{55.06} \\
 &  & SoftPaint & 1.0 & \textbf{0.099} & \textbf{21.51} & \textbf{0.060} & \textbf{0.245} & 54.60 \\
\midrule
\multirow{6}{*}{CogVideoX}
 & \multirow{2}{*}{\shortstack{Weak Edit\\(CLIP-T: 20.6$\pm$0.5)}} & DD        & 0.4 & 0.046 & 27.83 & 0.022 & 0.113 & \textbf{56.59} \\
 &  & SoftPaint & 0.2 & \textbf{0.043} & \textbf{29.15} & \textbf{0.019} & \textbf{0.112} & 56.39 \\
\cmidrule(lr){3-9}
 & \multirow{2}{*}{\shortstack{Medium Edit\\(CLIP-T: 22.4$\pm$0.5)}} & DD        & 0.6 & 0.059 & 25.15 & 0.033 & \textbf{0.171} & 56.79 \\
 &  & SoftPaint & 0.5 & \textbf{0.056} & \textbf{26.59} & \textbf{0.030} & \textbf{0.171} & \textbf{57.13} \\
\cmidrule(lr){3-9}
 & \multirow{2}{*}{\shortstack{Strong Edit\\(CLIP-T: 24.7$\pm$0.5)}} & DD        & 1.0 & 0.101 & 20.05 & 0.066 & 0.287 & 52.54 \\
 &  & SoftPaint & 0.9 & \textbf{0.089} & \textbf{21.57} & \textbf{0.058} & \textbf{0.250} & \textbf{54.53} \\
\midrule
\bottomrule
\end{tabular}
\end{table*}

\begin{table*}[t]
\centering
\small
\setlength{\tabcolsep}{4pt}
\caption{\textbf{Image editing at matched NFE on Qwen-Image (FlowEdit).} SoftPaint uses $20$ outer steps and two additional inner evaluations per step ($60$ NFE); DD uses $20$ or $60$ outer steps. Rows within each block are selected at the same CLIP-T anchor. The DD $20$-NFE and SoftPaint $60$-NFE rows reproduce the corresponding values in Table~\ref{tab:exp_image_main}.}
\label{tab:exp_image_equal_nfe}
\begin{tabular}{l l c c c c c c}
\toprule
Anchor & Method & NFE & LPIPS$_{\text{edit}}\downarrow$ & PSNR$_{\text{edit}}\uparrow$ & PSNR$_{\text{keep}}\uparrow$ & CLIP-IQA$\uparrow$ & FID$\downarrow$ \\
\midrule
\multirow{3}{*}{\shortstack{Weak Edit\\(CLIP-T: 30.1$\pm$0.2)}}
 & DD & 20 & 0.082 & 25.34 & 65.54 & \textbf{0.626} & 56.3 \\
 & DD & 60 & 0.088 & 25.19 & 65.50 & 0.620 & 54.33 \\
 & SoftPaint & 60 & \textbf{0.075} & \textbf{26.36} & \textbf{66.31} & \textbf{0.626} & \textbf{52.7} \\
\midrule
\multirow{3}{*}{\shortstack{Medium Edit\\(CLIP-T: 33.0$\pm$0.2)}}
 & DD & 20 & 0.131 & 21.31 & 65.34 & 0.619 & 102.2 \\
 & DD & 60 & 0.134 & 20.80 & 65.25 & 0.613 & 106.4 \\
 & SoftPaint & 60 & \textbf{0.121} & \textbf{22.05} & \textbf{66.29} & \textbf{0.627} & \textbf{99.9} \\
\midrule
\multirow{3}{*}{\shortstack{Strong Edit\\(CLIP-T: 35.5$\pm$0.2)}}
 & DD & 20 & 0.225 & 16.10 & 64.95 & 0.584 & 163.2 \\
 & DD & 60 & 0.224 & 15.88 & 64.94 & 0.581 & 160.01 \\
 & SoftPaint & 60 & \textbf{0.214} & \textbf{16.28} & \textbf{65.98} & \textbf{0.597} & \textbf{157.2} \\
\bottomrule
\end{tabular}
\end{table*}

\begin{table*}[h!]
\centering
\small
\setlength{\tabcolsep}{4pt}
\caption{\textbf{Image-editing on Qwen-Image at matched CLIP-T.}
Anchors are SoftPaint at $r\in\{0.2,0.6,1.0\}$. For DD and Naive Mixing we select the nearest measured $r$ within $\pm0.2$ CLIP-T of each anchor. Source is listed for reference. Naive Mixing has no match at the medium anchor.}
\label{tab:exp_image_main}
\begin{tabular}{l l c c c c c c}
\toprule
Anchor & Method & $r$ used & LPIPS$_{\text{edit}}\!\downarrow$ & PSNR$_{\text{edit}}\!\uparrow$ & PSNR$_{\text{keep}}\!\uparrow$ & CLIP-IQA$\uparrow$ & FID$\downarrow$ \\
\midrule
\multicolumn{8}{c}{\textit{Reference Points}} \\
\midrule
NA & Source            & 0  & 0.000 & 80.00 & 80.00 & 0.650 & 0.0 \\
NA & QwenImage-Edit   & NA & 0.306 & 16.19 & 25.39 & 0.667 & 129.5 \\
\midrule
\multicolumn{8}{c}{\textit{Soft-Mask Methods}} \\
\midrule
\multirow{3}{*}{\shortstack{Weak Edit\\(CLIP-T: 30.1$\pm$0.2)}}
 & Naive Mixing  & 0.8   & 0.057 & 28.92 & 65.58 & 0.595 & 46.2 \\
 & DD            & 0.4   & 0.082 & 25.34 & 65.54 & 0.626 & 56.3 \\
 & \textbf{SoftPaint} & \textbf{0.2} & \textbf{0.075} & \textbf{26.36} & \textbf{66.31} & \textbf{0.626} & \textbf{52.7} \\
\midrule
\multirow{3}{*}{\shortstack{Medium Edit\\(CLIP-T: 33.0$\pm$0.2)}}
 & Naive Mixing  & ---   & ---   & ---   & ---   & ---   & --- \\
 & DD            & 0.7   & 0.131 & 21.31 & 65.34 & 0.619 & 102.2 \\
 & \textbf{SoftPaint} & \textbf{0.6} & \textbf{0.121} & \textbf{22.05} & \textbf{66.29} & \textbf{0.627} & \textbf{99.9} \\
\midrule
\multirow{3}{*}{\shortstack{Strong Edit\\(CLIP-T: 35.5$\pm$0.2)}}
 & Naive Mixing  & 1.0   & 0.225 & 16.09 & 64.98 & 0.580 & 164.1 \\
 & DD            & 1.0   & 0.225 & 16.10 & 64.95 & 0.584 & 163.2 \\
 & \textbf{SoftPaint} & \textbf{1.0} & \textbf{0.214} & \textbf{16.28} & \textbf{65.98} & \textbf{0.597} & \textbf{157.2} \\
\bottomrule
\end{tabular}
\end{table*}


\section{Derivation of the DDL Update}
\label{app:ddl-derivation}

This section derives the damped exponential-integrator update used in the main text. Starting from the Langevin dynamics in Eq.~\eqref{eq:langevin_main}, a naive Euler--Maruyama discretization requires very small step sizes because the UR score contains a stiff linear component. Expanding Eq.~\eqref{eq:g_lambda} and expressing the model score via Tweedie's formula,
\[
\mathbf{s}_{\text{model}}(\mathbf{z},t) = \frac{\sqrt{\bar{\alpha}_t}\,\hat{\mathbf{x}}_0(\mathbf{z},t) - \mathbf{z}}{1-\bar{\alpha}_t},
\]
gives
\begin{equation}
\mathbf{g}(\mathbf{z},t) = \underbrace{-\frac{\mathbf{1}}{1-\bar{\alpha}_t}\,\mathbf{z}}_{\mbox{linear term}} \;+\; \underbrace{\frac{\sqrt{\bar{\alpha}_t}}{1-\bar{\alpha}_t}\Bigl(\boldsymbol{\gamma}_t\mathbf{x}_0^{\text{ref}} + (\mathbf{1}-\boldsymbol{\gamma}_t)\,\hat{\mathbf{x}}_0(\mathbf{z},t)\Bigr)}_{\mbox{nonlinear remainder}},
\end{equation}
where $\hat{\mathbf{x}}_0$ is the model's denoised estimate. The coefficient $(1-\bar{\alpha}_t)^{-1}$ becomes large when $1-\bar{\alpha}_t$ is small, which makes explicit Euler steps numerically fragile.

To stabilise the inner-loop sampler, we split the drift into a linear damping term and a remainder:
\begin{equation}
\mathbf{g}(\mathbf{z},t) = -A_t \mathbf{z} + \mathbf{C}_t(\mathbf{z}), \qquad
\mathbf{C}_t(\mathbf{z}) := \mathbf{g}(\mathbf{z},t) + A_t\mathbf{z}.
\end{equation}
In practice, we choose
\[
A_t = \frac{1+\lambda\beta_t(\mathbf{r})}{1-\bar{\alpha}_t},
\]
so that the damping strength increases with the guidance factor $\lambda\beta_t(\mathbf{r})$ and remains stable even when $\lambda$ is large. This damping rule is a simplified variant of the damped Langevin dynamics introduced in LanPaint~\cite{zheng2025lanpaint}.

If we treat $\mathbf{C}_t(\mathbf{z})$ as approximately constant over a short interval $\Delta\tau$, the SDE becomes an Ornstein--Uhlenbeck process with a closed-form transition, yielding Eq.~\eqref{eq:ou_update}. This update uses one score evaluation per step and remains stable for arbitrarily large $\Delta\tau$. In the limit $\Delta\tau\to\infty$, it relaxes to the stationary Gaussian
\[
\mathcal{N}(A_t^{-1}\mathbf{C}_t,\,A_t^{-1}\mathbf{I}).
\]
The corresponding limiting variance is
\[
A_t^{-1}\mathbf{I} = \frac{1-\bar{\alpha}_t}{1+\lambda\beta_t(\mathbf{r})}\mathbf{I} \;\preceq\; (1-\bar{\alpha}_t)\mathbf{I},
\]
which is bounded above by the natural diffusion variance. Therefore, the sampling trajectory remains contained even under strong guidance, unlike standard Langevin updates with overly large explicit step sizes.

\section{Ablation Studies}
\label{app:ablations}

For completeness, the main video experiments use the official default denoising schedules of each backbone: $50$ steps for CogVideoX-5B and $40$ steps for Wan2.2.

We ablate SoftPaint's two inner-loop hyperparameters---Langevin step size $s$ and number of inner updates $n$---on the Wan2.2 backbone over DAVIS-Edit, averaging all metrics across $r\in\{0.2,0.4,0.6,0.8,1.0\}$ to avoid cherry-picking a single operating point. We use these same defaults ($n{=}2$, $s{=}0.2$) for image editing without any per-domain re-tuning.

\subsection{Inner-loop hyperparameters}
\label{app:abl-ns}

\begin{figure*}[h!]
\centering
\includegraphics[width=\textwidth]{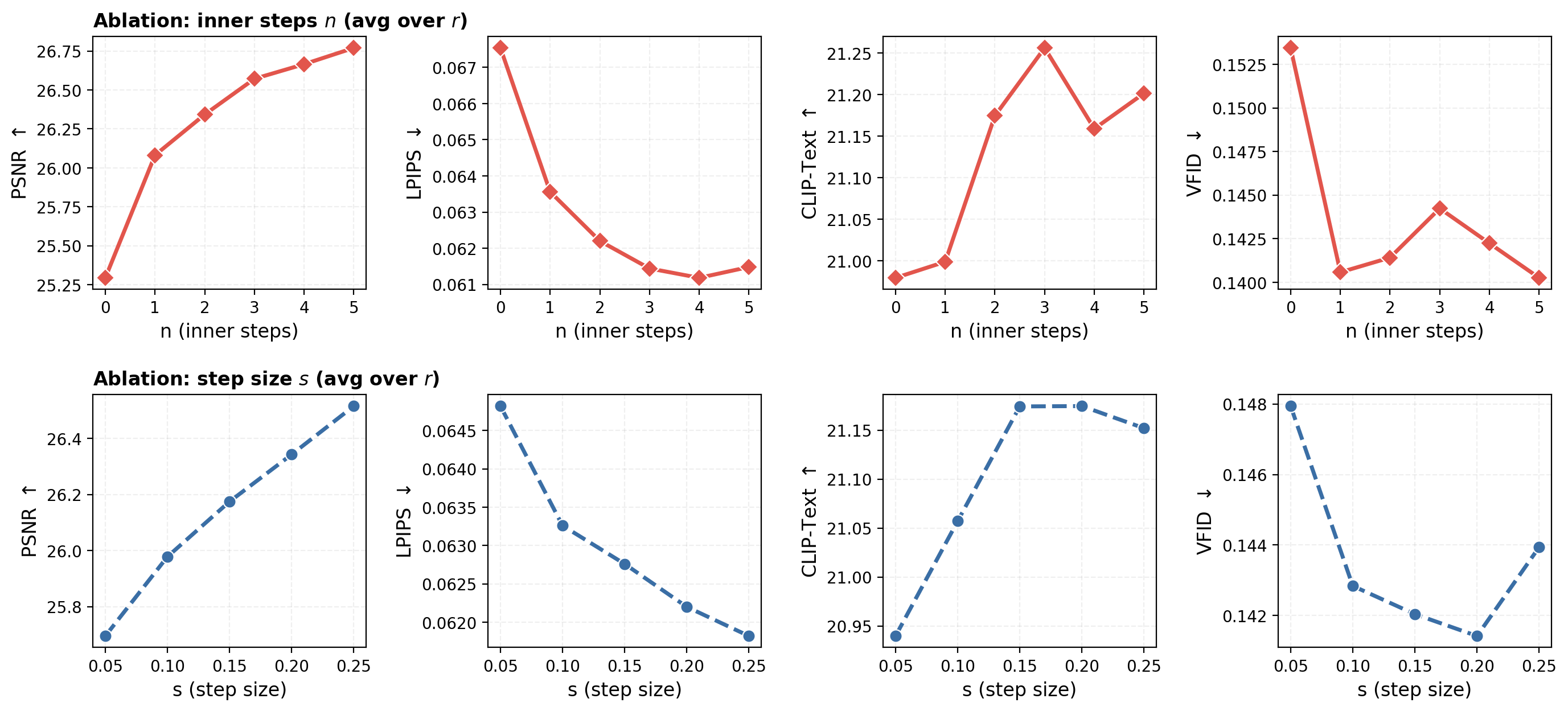}
\caption{\textbf{Ablation of $n$ and $s$ on soft video editing task(DAVIS-Edit) under backbone Wan2.2.} Top: $n$ varied with $s{=}0.2$ fixed. Bottom: $s$ varied with $n{=}2$ fixed. All metrics averaged over $r\in\{0.2,0.4,0.6,0.8,1.0\}$. $n{=}2$, $s{=}0.2$ are the default operating point, also used for image editing without re-tuning.}
\label{fig:abl_video_ns}
\end{figure*}

Figure~\ref{fig:abl_video_ns} reports PSNR, LPIPS, CLIP-Text, and VFID against $n$ (top row, $s{=}0.2$ fixed) and $s$ (bottom row, $n{=}2$ fixed). Increasing $n$ from 0 to 2 improves PSNR by $+1.05$ and lowers LPIPS by $0.0053$; $n{=}3{\to}5$ adds only marginal gains ($+0.43$ PSNR) while raising the inner-loop cost $2.5{\times}$. Similarly, increasing $s$ from $0.05$ to $0.2$ yields $+0.65$ PSNR with diminishing returns beyond $s{=}0.2$. The sweet spot $n{=}2$, $s{=}0.2$ balances quality and compute---a single Langevin step ($n{=}1$) is too noisy, while larger budgets bring negligible improvement.

\subsection{Runtime}
\label{app:runtime}

SoftPaint's runtime is dominated by the inner-loop budget $n$, since each inner DDL update reuses the backbone denoiser. On Qwen-Image (roughly 20B MMDiT) at $1024\times1024$ with $20$ diffusion steps on a single accelerator with approximately $65$ TFLOPS of FP16 compute, measuring only the CUDA-synchronized denoising loop (excluding model loading, encoder passes, VAE decoding, and post-processing), DD averages $10.2$~s/image while SoftPaint with $n{=}2$ averages $29.4$~s/image. SoftPaint uses the same peak GPU memory as standard forward-only backbone inference, i.e., it does not introduce extra memory beyond ordinary inference.

\subsection{Redo-aware warm start}
\label{app:abl-warm-start}

Before running the inner Langevin loop, we optionally initialise the chain from a redo-aware warm start instead of the raw model sample $\mathbf{x}_t$. Recall from Eq.~\eqref{eq:gamma_t_final} that $\gamma_t^{(i)} > 1$ corresponds to the extrapolation regime, where the reference constraint is amplified. For such pixels, it is beneficial to initialise directly from the noised reference
\[
\mathbf{x}_t^{\text{ref}} = \sqrt{\bar{\alpha}_t}\mathbf{x}_0^{\text{ref}} + \sqrt{1-\bar{\alpha}_t}\boldsymbol{\epsilon}
\]
rather than from $\mathbf{x}_t$. Concretely, we define a binary mask
\[
\mathbf{m} = \mathbb{I}[\boldsymbol{\gamma}_t(\mathbf{r}) > 1]
\]
and set
\[
\tilde{\mathbf{x}}_t = (\mathbf{1}-\mathbf{m})\odot\mathbf{x}_t + \mathbf{m}\odot\mathbf{x}_t^{\text{ref}}.
\]
This warm start does \emph{not} alter the stationary distribution of the inner Langevin chain; its sole purpose is to shorten burn-in. Without it, the chain initialises at a low-likelihood point of $\pi(\mathbf{x}_t)$ and, at our small budget of $n{=}2$, leaves residual noise at mask boundaries. 

Table~\ref{tab:cogvideox_warmup_ablation} isolates the $r{=}0.8$ case, where the difference between the three configurations is especially clear. Naive mixing stays highly similar to the source and has very low CLIP-T, indicating almost no editing ability at this mask strength. Adding the warm start substantially improves editability, as reflected by the large CLIP-T increase. SoftPaint further adds the UR score and the DDL Langevin inner loop, improving editability again while also improving LPIPS and VFID relative to the warm-start-only variant.

\begin{table*}[t]
\centering
\small
\setlength{\tabcolsep}{5pt}
\caption{\textbf{Warm-start ablation for soft video editing at $r=0.8$ on the CogVideoX backbone.} Naive mixing has almost no editing ability at this mask strength. Adding the warm start restores editing, and SoftPaint further improves editability with the UR score and DDL Langevin inner loop while improving LPIPS and VFID relative to warm start alone. All rows use 50 denoising steps.}
\label{tab:cogvideox_warmup_ablation}
\begin{tabular}{l c c c c}
\toprule
Method & PSNR$\uparrow$ & LPIPS$\downarrow$ & CLIP-T$\uparrow$ & VFID$\downarrow$ \\
\midrule
Naive mix. & 32.46 & 0.0333 & \makecell[l]{19.95\\\textcolor{gray}{\scriptsize almost no edit}} & 0.0668 \\
Naive mix.\ + warm start & 22.22 & 0.0786 & \makecell[l]{24.10\\\textcolor{ForestGreen}{\scriptsize $\uparrow$ edit}} & 0.2410 \\
SoftPaint & 23.15 & 0.0757 & \makecell[l]{24.32\\\textcolor{ForestGreen}{\scriptsize $\uparrow$ edit}} & 0.2305 \\
\bottomrule
\end{tabular}
\\[2pt]
\parbox{0.82\textwidth}{\scriptsize \textit{Note:} Lower PSNR here mainly reflects larger deviation from the source under stronger edits, rather than worse perceptual quality.}
\end{table*}

\subsection{Binary-mask inpainting: isolating the inner-loop operator}
\label{app:binary-inpainting}

We also test whether the DDL update offers a benefit beyond repeated denoise--replace--renoise updates under binary masks. On the ImageNet binary-inpainting benchmark, we use $1{,}000$ validation images at $256\times256$ resolution, an ADM diffusion model~\citep{Dhariwal2021DiffusionMB}, $20$ outer Euler steps, and box or checkerboard masks. The comparison holds the outer schedule and binary mask fixed. Replace~\citep{song2019generative} uses no inner loop; the RePaint-style~\citep{lugmayr2022repaint} and SoftPaint variants use the same number of inner iterations and matched model-evaluation budgets at each $n$.

Table~\ref{tab:binary-inpainting} reports LPIPS and FID for $n\in\{5,10\}$. SoftPaint yields lower values than the RePaint-style inner loop on both masks at both budgets. The comparison isolates the gain from DDL in the binary-mask setting, where spatially varying soft-mask weights are absent. Replace is included as a $20$-step reference and does not have the same NFE as the two iterative variants.

\begin{table}[H]
\centering
\small
\setlength{\tabcolsep}{7pt}
\caption{\textbf{Binary inpainting on ImageNet with ADM.} LPIPS $\downarrow$ / FID $\downarrow$ on $1{,}000$ validation images at $256\times256$. All methods use $20$ outer Euler steps. RePaint-style and SoftPaint have matched inner-loop budgets for each $n$; Replace has no inner loop.}
\label{tab:binary-inpainting}
\begin{tabular}{l c c c c}
\toprule
Mask & Inner steps $n$ & Replace~\citep{song2019generative} & RePaint-style~\citep{lugmayr2022repaint} & SoftPaint \\
\midrule
\multirow{2}{*}{Box}
 & 5  & \multirow{2}{*}{0.229 / 75.7} & 0.216 / 62.8 & \textbf{0.174 / 48.1} \\
 & 10 &                            & 0.215 / 61.0 & \textbf{0.165 / 43.5} \\
\midrule
\multirow{2}{*}{Checkerboard}
 & 5  & \multirow{2}{*}{0.406 / 146.4} & 0.137 / 31.4 & \textbf{0.118 / 21.6} \\
 & 10 &                             & 0.135 / 32.8 & \textbf{0.110 / 19.9} \\
\bottomrule
\end{tabular}
\end{table}

\subsection{Choice of exponent $\eta$ in $\Sigma(\mathbf{r};\eta)$}
\label{app:abl-eta}

\begin{figure}[h!]
\centering
\includegraphics[width=0.4\columnwidth]{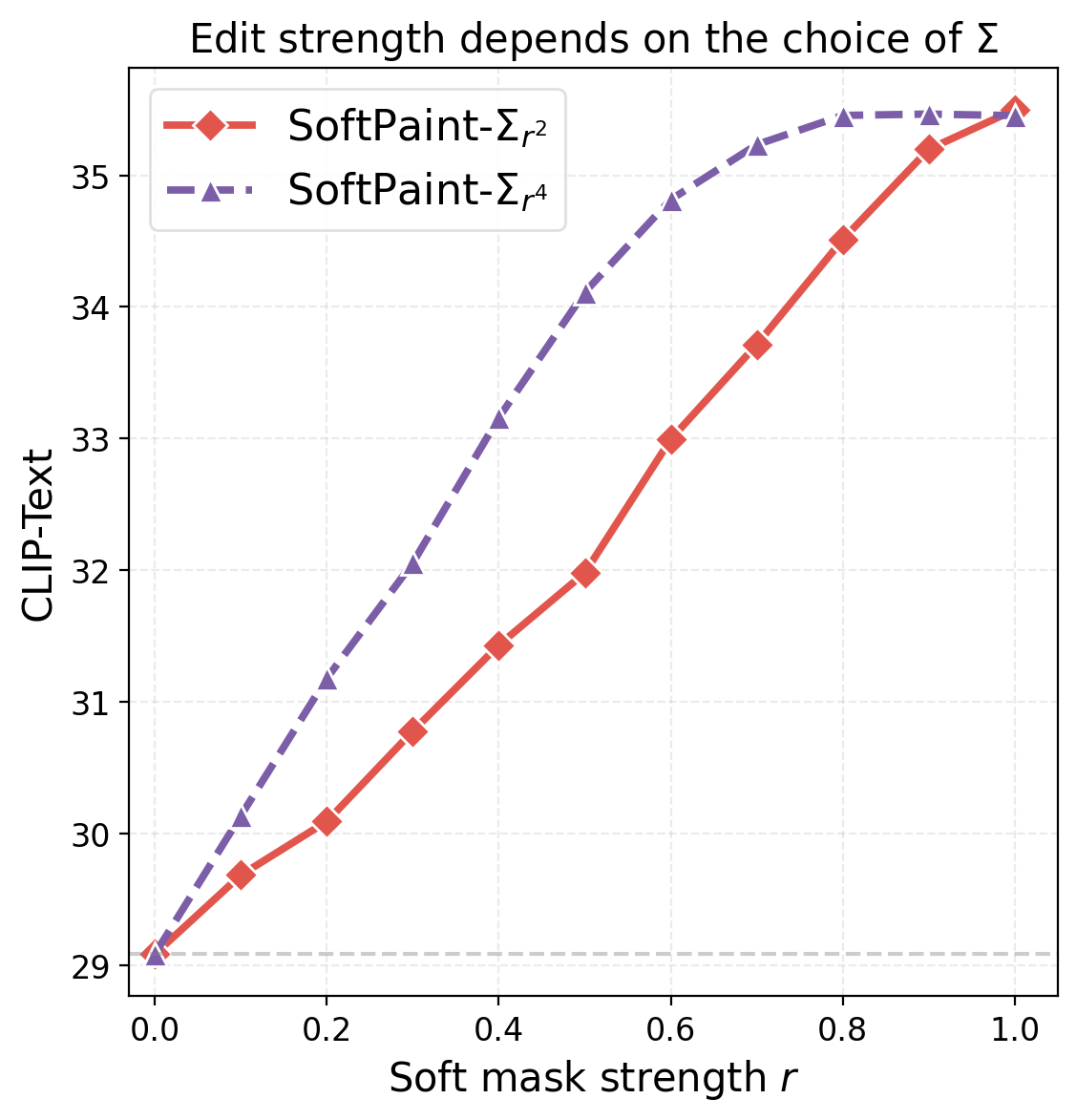}
\caption{Effect of the exponent $\eta$ in $\Sigma(\mathbf{r};\eta)$ on edit strength, measured on the Flow Edit Benchmark with the Qwen-Image model. The two curves correspond to $\eta=2$ and $\eta=4$, showing how the choice of exponent changes the effective redo strength for each mask value $r$, and hence $\mathrm{clip}_t$.}
\label{fig:sigma_choice}
\end{figure}

We also ablate the exponent $\eta$ in $\Sigma(\mathbf{r};\eta) = (1+2\lambda)\,\frac{1-(1-\mathbf{r})^{\eta}}{(1-\mathbf{r})^{\eta}+\epsilon}$. Figure~\ref{fig:sigma_choice} compares $\eta=2$ and $\eta=4$ on the Flow Edit Benchmark with the Qwen-Image model. We choose $\eta=2$ because it makes $\mathrm{clip}_t$ approximately linear in $r$, yielding a more predictable mapping from mask value to redo strength.

\section{Derivation of the Expectation Form for the Reference Factor} \label{app:ur-derivation}

We start from the forward transition density
\begin{equation}
p(x_t^{(i)}\mid x_0^{(i)}) = \mathcal{N}\!\bigl(x_t^{(i)}\mid \sqrt{\bar{\alpha}_t}\,x_0^{(i)},\,1-\bar{\alpha}_t\bigr)
\propto \exp\!\left(-\frac{(x_t^{(i)}-\sqrt{\bar{\alpha}_t}\,x_0^{(i)})^2}{2(1-\bar{\alpha}_t)}\right).
\end{equation}
Raising this density to the power $\gamma_t^{(i)}$ and absorbing constants into proportionality gives
\begin{equation}
p(x_t^{(i)}\mid x_0^{(i)})^{\gamma_t^{(i)}}
\propto \exp\!\left(-\frac{\gamma_t^{(i)}(x_t^{(i)}-\sqrt{\bar{\alpha}_t}\,x_0^{(i)})^2}{2(1-\bar{\alpha}_t)}\right).
\end{equation}
We choose
\begin{equation}
\gamma_t^{(i)} = (1+\lambda)\,\beta_t^{(i)}(\mathbf{r}),\qquad
\beta_t^{(i)}(\mathbf{r}) = \frac{1-\bar{\alpha}_t}{1-\bar{\alpha}_t + \bar{\alpha}_t\,\Sigma^{(i)}(\mathbf{r})}.
\end{equation}
Substituting $\gamma_t^{(i)}$ into the exponent yields
\begin{align}
p(x_t^{(i)}\mid x_0^{(i)})^{\gamma_t^{(i)}}
&\propto \exp\!\left(-\frac{(1+\lambda)\,\beta_t^{(i)}(\mathbf{r})\,(x_t^{(i)}-\sqrt{\bar{\alpha}_t}\,x_0^{(i)})^2}{2(1-\bar{\alpha}_t)}\right) \notag \\
&= \exp\!\left(-\frac{1+\lambda}{2\bigl(1-\bar{\alpha}_t + \bar{\alpha}_t\,\Sigma^{(i)}(\mathbf{r})\bigr)}\bigl(x_t^{(i)}-\sqrt{\bar{\alpha}_t}\,x_0^{(i)}\bigr)^2\right). \label{eq:lhs_exp}
\end{align}

Now consider the expectation under a Gaussian smoothing distribution
$\tilde{x}_0^{(i)}\sim\mathcal{N}\bigl(x_0^{(i)},\Sigma^{(i)}(\mathbf{r})\bigr)$:
\begin{equation}
\mathbb{E}_{\tilde{x}_0^{(i)}}\bigl[p(x_t^{(i)}\mid\tilde{x}_0^{(i)})\bigr]
\propto \int \exp\!\left(-\frac{(x_t^{(i)}-\sqrt{\bar{\alpha}_t}\,\tilde{x}_0^{(i)})^2}{2(1-\bar{\alpha}_t)}\right)
\exp\!\left(-\frac{(\tilde{x}_0^{(i)}-x_0^{(i)})^2}{2\Sigma^{(i)}(\mathbf{r})}\right) d\tilde{x}_0^{(i)}.
\end{equation}
The integrand is a product of two Gaussian exponents. Completing the square in $\tilde{x}_0^{(i)}$ and integrating out the variable leaves an exponent that depends only on $x_t^{(i)}$ and $x_0^{(i)}$. The result (up to a constant factor) is
\begin{equation}
\mathbb{E}_{\tilde{x}_0^{(i)}}\bigl[p(x_t^{(i)}\mid\tilde{x}_0^{(i)})\bigr]
\propto \exp\!\left(-\frac{(x_t^{(i)}-\sqrt{\bar{\alpha}_t}\,x_0^{(i)})^2}{2\bigl(1-\bar{\alpha}_t + \bar{\alpha}_t\,\Sigma^{(i)}(\mathbf{r})\bigr)}\right).
\end{equation}
Raising this expectation to the power $1+\lambda$ gives
\begin{equation}
\Bigl[\mathbb{E}_{\tilde{x}_0^{(i)}}\,p(x_t^{(i)}\mid\tilde{x}_0^{(i)})\Bigr]^{1+\lambda}
\propto \exp\!\left(-\frac{1+\lambda}{2\bigl(1-\bar{\alpha}_t + \bar{\alpha}_t\,\Sigma^{(i)}(\mathbf{r})\bigr)}\bigl(x_t^{(i)}-\sqrt{\bar{\alpha}_t}\,x_0^{(i)}\bigr)^2\right). \label{eq:rhs_exp}
\end{equation}

Comparing \eqref{eq:lhs_exp} and \eqref{eq:rhs_exp}, we observe that the two expressions share exactly the same functional form in $x_t^{(i)}$. Since normalising constants do not affect the reverse sampling dynamics, we conclude
\begin{equation}
p\!\left(x_t^{(i)}\mid x_0^{(i)}\right)^{\gamma^{(i)}_t}
\propto \Bigl[\mathbb{E}_{\tilde{x}_0^{(i)}\sim\mathcal{N}(x_0^{(i)},\Sigma^{(i)}(\mathbf{r}))}\,
p(x_t^{(i)}\mid \tilde{x}_0^{(i)})\Bigr]^{1+\lambda},
\end{equation}
which justifies the choice of $\gamma^{(i)}_t$ stated in the main text.

\section{Additional Qualitative Results}
\label{app:qualitative}

\begin{figure}[h]
\centering
\includegraphics[width=\columnwidth]{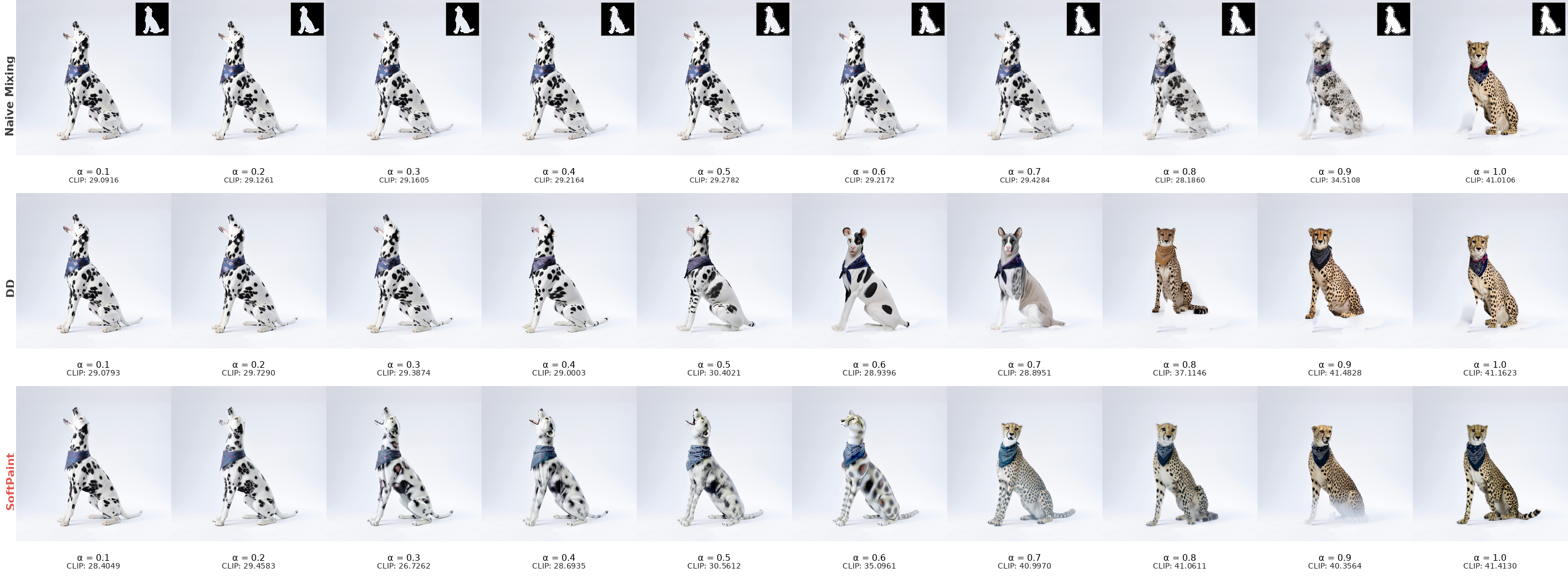}
\caption{\textbf{Soft-mask image editing: ``Dalmatian'' $\to$ ``Cheetah''.} Source: ``A Dalmatian dog sitting on a white background\ldots'' Target: ``A cheetah sitting on a white background\ldots, looking to the camera.'' Top: Naive Mixing. Middle: DD. Bottom: SoftPaint (ours). $r\in\{0.1,\ldots,1.0\}$ left to right; mask inset at top-right.}
\label{fig:app_qual_samples_1}
\end{figure}

\begin{figure}[h]
\centering
\includegraphics[width=\columnwidth]{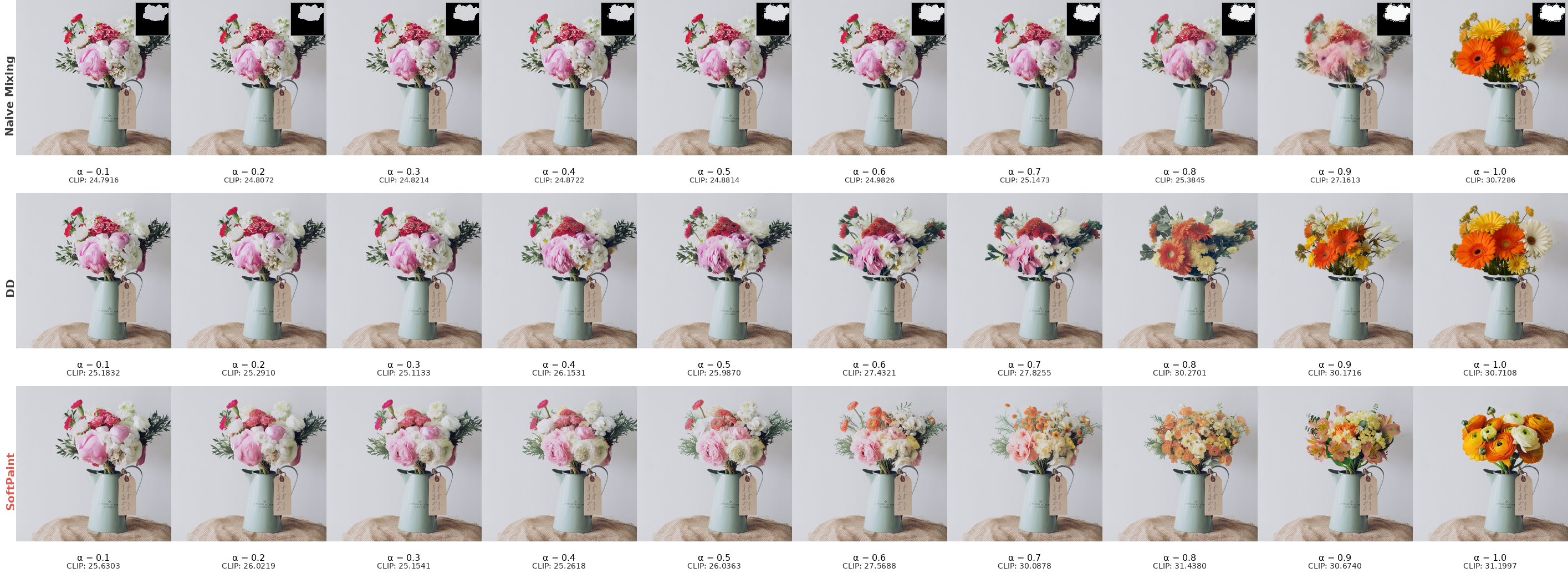}
\caption{\textbf{Soft-mask image editing: ``Pink Flowers'' $\to$ ``Orange Flowers''.} Source: ``A vase filled with a beautiful bouquet of pink, red and white flowers.'' Target: ``\ldots orange, yellow and white flowers.'' Layout as in Fig.~\ref{fig:app_qual_samples_1}.}
\label{fig:app_qual_samples_2}
\end{figure}

\begin{figure*}[t]
\centering
\includegraphics[width=\textwidth]{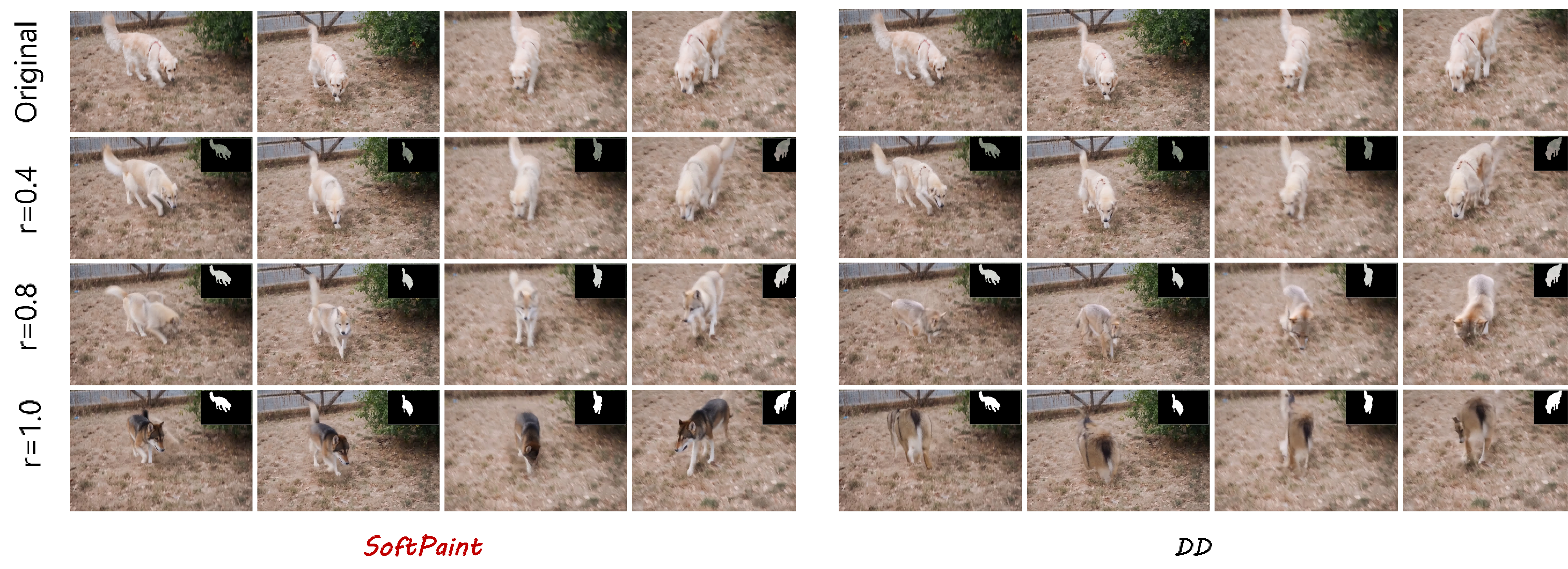}
\caption{\textbf{Soft-mask video editing: ``Dog'' $\to$ ``Wolf''} at three soft-mask strengths $r$ on CogVideoX. Top row: original DAVIS-Edit frames ($r{=}0$, no edit). Subsequent rows: SoftPaint and DD outputs at $r{=}0.4$, $0.8$, $1.0$.}
\end{figure*}

\end{document}